\documentclass[twocolumn]{aastex63}

\usepackage{bm}
\usepackage{hyperref}
\usepackage{breakcites}
\usepackage{amsmath,amssymb,amsthm} 
\usepackage{url} 
\usepackage{float}
\usepackage{graphicx}
\usepackage{subfigure}
\usepackage{color}
\usepackage{lineno}

\usepackage{chngcntr}
\newcommand{\eqb}{\begin{eqnarray}}
\newcommand{\eqe}{\end{eqnarray}}
\newcommand{\bi}{\begin{itemize}}
\newcommand{\ei}{\end{itemize}}

\newcommand{\txs}{TXS~0506+056}
\newcommand{\icv}{IceCube-200107A}
\newcommand{\hsp}{3HSP~J095507.9+355101}
\newcommand{\nustar}{\emph{NuSTAR}}
\newcommand{\nicer}{\emph{NICER}}
\newcommand{\swift}{\emph{Swift}}
\newcommand{\fermi}{\emph{Fermi}}

\newcommand{\eFenu}{\mbox{$\varepsilon_\nu F_{\varepsilon_\nu}$}}

\newcommand{\sth}{\sigma_{\rm T}}
\newcommand{\fpg}{f_{\rm mes}}
\newcommand{\dop}{\mathcal{D}}
\newcommand{\ergcmsqs}{$\mathrm{erg\,cm^{-2}\,s^{-1}}$}
\newcommand{\ergs}{$\mathrm{erg\,s^{-1}}$}
 
\received{14 May 2020}
\accepted{21 July 2020}
\submitjournal{ApJ}

\shorttitle{Multimessenger modeling of an extreme blazar}
\shortauthors{Petropoulou et al.}

\begin{document}

\title{Comprehensive Multimessenger Modeling of the Extreme Blazar \hsp \, and Predictions for IceCube}

\author[0000-0001-6640-0179]{Maria Petropoulou}
\affil{Department of Astrophysical Sciences, Princeton University,
Princeton, NJ 08544, USA}
 
\author[0000-0002-0525-3758]{Foteini Oikonomou}
\affil{European Southern Observatory, Karl-Schwarzschild-Str. 2, Garching bei M{\"u}nchen D-85748, Germany}
\affil{Technische Universit{\"a}t M{\"u}nchen, Physik-Department, 
James-Frank-Str. 1, D-85748 Garching bei M{\"u}nchen, Germany}
\affil{Institutt for fysikk, NTNU, Trondheim, Norway} 

\author{Apostolos Mastichiadis}
\affil{Department of Physics, National and Kapodistrian University of Athens, Panepistimiopolis, GR 15783 Zografos, Greece}

\author[0000-0002-5358-5642]{Kohta~Murase}
\affil{Department of Physics, Pennsylvania State University,
  University Park, PA 16802, USA}
\affil{Department of Astronomy \& Astrophysics,
  Pennsylvania State University,
  University Park, PA 16802, USA}
\affil{Center for Particle \& Gravitational Astrophysics,
  Institute for Gravitation and the Cosmos,
  Pennsylvania State University,
  University~Park, PA 16802, USA}
\affil{Center for Gravitational Physics, Yukawa Institute for Theoretical Physics, Kyoto, Kyoto 606-8502 Japan}

\author[0000-0002-4707-6841]{Paolo Padovani}
\affil{European Southern Observatory, Karl-Schwarzschild-Str. 2, Garching bei M{\"u}nchen D-85748, Germany}
\affil{Associated to INAF - Osservatorio Astronomico di Roma, via Frascati 33, I-00040 Monteporzio Catone, Italy}

\author[0000-0003-3902-3915]{Georgios Vasilopoulos}
\affil{Department of Astronomy, Yale University, New Haven, CT 06520-8101, USA}

\author[0000-0002-2265-5003]{Paolo Giommi}
\affil{Institute for Advanced Study, Technische Universit{\"a}t M{\"u}nchen, Lichtenbergstrasse 2a, D-85748 Garching bei M{\"u}nchen, Germany}
\affil{Associated  to Agenzia Spaziale Italiana, ASI, via del Politecnico s.n.c., I-00133 Roma Italy
}
\affil{ICRANet, Piazzale della Repubblica 10, I-65122, Pescara, Italy
}

\begin{abstract}
3HSP~J095507.9+355101 is an extreme blazar which has been possibly associated with a high-energy neutrino (IceCube-200107A) detected one day before the blazar was found to undergo a hard X-ray flare. We perform a comprehensive study of the predicted multimessenger emission from 3HSP~J095507.9+355101 during its recent X-ray flare, but also in the long term. We focus on one-zone leptohadronic models, but we also explore alternative scenarios: (i) a blazar-core model, which considers neutrino production in the inner jet, close to the supermassive black hole; (ii) a hidden external-photon model, which considers neutrino production in the jet through interactions with photons from a weak broad line region; (iii) a proton synchrotron model, where high-energy protons in the jet produce $\gamma$-rays via synchrotron; and (iv) an intergalactic cascade scenario, where neutrinos are produced in the intergalactic medium by interactions of a high-energy cosmic-ray beam escaping the jet. The Poisson probability to detect one muon neutrino in ten years from 3HSP~J095507.9+355101 with the real-time IceCube alert analysis  
is $\sim 1\%$ ($3\%$) for the most optimistic one-zone leptohadronic model (the multi-zone blazar-core model). Meanwhile, detection of one neutrino during the  44-day-long high X-ray flux-state period following the neutrino detection is 0.06\%, according to our most optimistic leptohadronic model. The most promising scenarios for neutrino production also predict strong  intra-source $\gamma$-ray attenuation above $\sim100$~GeV. If the association is real, then IceCube-Gen2 and other future detectors should be able to provide additional evidence for neutrino production in 3HSP~J095507.9+355101 and other extreme blazars.
\end{abstract}


\section{Introduction} \label{sec:intro}
The IceCube Neutrino Observatory\footnote{\url{http://icecube.wisc.edu}} 
reported the observation of neutrinos of 
astrophysical origin in 2013~\citep{PhysRevLett.111.021103,icecubeScience,PhysRevLett.113.101101}. Updated analyses since then have strengthened the significance of the observation~\citep{2019ICRC...36.1017S,Schneider19,2020arXiv200109520I}. 

In 2018 the IceCube Collaboration reported the observation of a
$\gtrsim 290$~TeV muon neutrino, IceCube-170922A, coincident with the peak of a $\sim6$-month-long $\gamma$-ray flare of the blazar TXS\,0506+056~\citep{IceCube:2018dnn}, whose redshift was later determined as $z = 0.3365$~\citep{Paiano:2018qeq}. Electromagnetic follow-up of the blazar led to a detection by several instruments, including MAGIC at energies exceeding 100~GeV. The correlation of the neutrino with the flare of TXS 0506+056 is
inconsistent with the hypothesis of arising by chance at the $3-3.5\sigma$ level. An archival search further revealed $13\pm5$  high-energy neutrinos in
the direction of TXS 0506+056 during a 6-month period in 2014-2015~\citep{IceCube:2018cha}. 
These events were not accompanied by a GeV $\gamma$-ray flare, and there was no evidence of enhanced flux at lower energies either \citep{IceCube:2018cha,Garrappa2019}. Such an accumulation of events is inconsistent
with arising from a background fluctuation at the 3.5$\sigma$ level. The results summarised above make \txs, an intermediate-peaked blazar (IBL)\footnote{Based on the
rest-frame frequency of the low-energy (synchrotron) hump, blazars are divided into low-energy peaked (LBL)
sources $(\nu_p<10^{14}$~Hz [$<$ 0.41 eV]), intermediate-energy peaked (IBL) sources
($10^{14}$~Hz~$<\nu_p< 10^{15}$~Hz [0.41 eV -- 4.1 eV]), and
high-energy  peaked (HBL) sources  ($\nu_p> 10^{15}$~Hz [$>$ 4.1 eV]) \citep{Padovani:1994sh,Abdo:2010nz}.}, the first astrophysical source to be associated with a high-energy neutrino at such significance. An additional indication of association of IBL and HBL sources with high-energy neutrinos has since been reported by~\cite{Giommi:2020hbx}; for an indication of association of high-energy neutrinos with blazars in general, see \cite{franckowiak2020patterns}.  

In January 2020 IceCube reported the observation of the high-energy neutrino, IceCube-200107A~\citep{2020GCN.26655....1I}. 
Electromagnetic follow-up of sources within the uncertainty region of the neutrino arrival direction led to the detection of an X-ray flare from the HBL blazar \hsp~\citep{2020GCN.26669....1G,2020ATel13394....1G,2020Atel13395....1K}, which
is part of the 3HSP catalogue~\citep{2019arXiv190908279C}. In fact, with a peak synchrotron frequency of $\nu_s \sim 5\times 10^{17}$~Hz, the source belongs to the rare class of extreme blazars~\citep{2001A&A...371..512C,Biteau2020}. It has also been detected by the {\it Fermi}-LAT as a $\gamma$-ray emitting source and is thus also included in the 4FGL catalog~\citep{2020ApJS..247...33A}. Subsequent to the detection of the X-ray flare, the redshift of the source was determined to be $z = 0.557$~\citep{2020MNRAS.495L.108P, Paliya2020}. 

Detailed observations of the source starting from the day following the IceCube alert were reported by some of us in~\citet{Giommi2020} (see also~\citealp{Paliya2020}). The chance probability of the observed association was estimated under several assumptions about the underlying source population in Section 3 of~\citet{Giommi2020}. An exact significance cannot, however, be established since these are {\it a posteriori} estimates. All in all, this is an interesting observation corroborating a trend of association between blazars and a fraction of IceCube neutrinos. Using analytical arguments, \cite{Giommi2020} estimated that \hsp \, might have produced at most $\sim 10^{-2}$ muon and anti-muon neutrinos during its recent flare, in line with the estimates for the 2017 flare of \txs~\citep[e.g.,][]{Keivani:2018rnh,Murase:2018iyl,Gao2019, Zhang:2019htg}.

In this work, we perform detailed multimessenger modeling of \hsp~to assess the expected neutrino emission of the source during its recent X-ray flare, and during the entire lifetime of IceCube operations. We focus 
primarily on the one-zone model for blazar emission, where neutrino and photon emissions are co-spatially produced in the blazar jet, but also discuss several alternative scenarios for neutrino production.  This is the first comprehensive study about the neutrino emission of an extreme blazar, and is motivated by the fact that \hsp~is the first extreme blazar to have been possibly associated with a high-energy neutrino. 

This paper is organised as follows. In Section~\ref{sec:model} we present the one-zone leptohadronic model used for the calculation of the neutrino emission of \hsp~and in Section~\ref{sec:code} we describe the adopted numerical approach. 
In Section~\ref{sec:results} we present the results of the  standard leptohadronic modeling of the X-ray flare of \hsp~after a brief description of the observational data (Section~\ref{sec:data}), and the model parameter selection (Section~\ref{sec:param}). We continue in Section~\ref{sec:longterm} with a presentation of the results for the long-term neutrino 
emission of 
the source. In Section~\ref{sec:other} we present alternative scenarios for neutrino production in \hsp, which include emission from the blazar  core, a hidden external photon model, a proton synchrotron emission model, and emission from an intergalactic cascade induced by a high-energy cosmic-ray beam escaping the blazar. In Section~\ref{sec:discussion} 
we discuss the implications of our model on the jet energetics, the relation between electromagnetic observations and expected neutrino flux, and the detection of \icv. We conclude in Section~\ref{sec:summary} with a brief summary of our results.

In this paper we adopt a cosmology with $\Omega_M=0.29$, $\Omega_\Lambda=0.71$,  and $H_0=69.6$~km s$^{-1}$ Mpc$^{-1}$ \citep{Bennett2014}. The redshift of \hsp \, 
corresponds to a luminosity distance $d_L\simeq3262$~Mpc.

\section{The one-zone leptohadronic model}\label{sec:model}
We adopt the standard one-zone leptohadronic model for blazar emission \citep[e.g.,][and references therein]{Petropoulou:2015upa,cerruti15}. According to this, the blazar (radiation) zone is approximated by a spherical blob of radius $R^\prime$ with magnetic field of strength $B^\prime$, moving towards the observer with a Doppler factor $\dop$. Henceforth, quantities measured in the co-moving frame of the blob are denoted with primes. Protons and electrons, which are accelerated by some mechanism into non-thermal energy distributions, are injected isotropically in the volume of the blob at a constant rate $\mathcal{Q}^{inj}_i$, which  translates to an injection luminosity $L^\prime_i$ (with $i=e,p$).
Particles are also assumed to escape on a timescale $t^\prime_{i,esc}$ which, for simplicity, is taken to be equal to the light-crossing time of the blob $R^\prime/c$ for both electrons and protons. The remaining free parameters of the one-zone leptohadronic model are related to the shape of the accelerated proton and electron energy spectra at injection. These will be discussed in the context of SED modeling in Section~\ref{sec:param}. 

Broadband non-thermal radiation is produced via a network of radiative processes involving charged particles, magnetic fields, and low-energy radiation, which can be produced by the particles themselves or/and can be unrelated to the particles (i.e., external to the blob). Relativistic protons lose energy by synchrotron radiation, photomeson production, and photopair (Bethe-Heitler) production. The last two processes, together with photon-photon pair production (i.e., electron-positron production by two photon annihilation), are an important source of secondary electron and positron pairs. The latter, same as the accelerated electrons (primary electrons), lose energy by synchrotron radiation and inverse Compton scattering. Photons are  therefore  produced in a variety of ways, namely synchrotron and Compton processes of primary electrons and secondary pairs, synchrotron radiation or protons and charged mesons, and decay of neutral pions. Photon-photon pair production, synchrotron self-absorption, and escape from the blob are processes that act as sinks of photons.  

The decay of charged pions leads to the production of high-energy muon and electron neutrinos\footnote{This term refers to both neutrinos and anti-neutrinos ($\nu+\bar{\nu}$).}, which escape the blob on a timescale $R^\prime/c$ without undergoing any interactions.  Neutrons, which are also a by-product of the photomeson production process~\citep[e.g.,][]{Kirk1989,Atoyan:2002gu,Dermer:2012rg,Murase:2018iyl,Zhang:2019htg}, can escape almost unimpeded from the radiation zone for typical parameters, as those used in this work (see e.g., Section~\ref{sec:param}). As long as the escaping protons and neutrons are energetic enough, they are susceptible to photomeson production interactions with ambient photons in the galactic and intergalactic space, such as the cosmic microwave and infrared backgrounds, producing additional high-energy neutrinos \citep{Stecker1973}. Neutrons also rapidly decay into protons \citep{Sikora1987,Kirk1989,Giovanoni1990, Atoyan:2001ey}, leading also to high-energy neutrino production. Our study focuses on the neutrino emission from the blazar zone. Hence, we do not consider additional contributions to the neutrino flux from escaping high-energy nucleons, till  Section~\ref{sec:cascade}, where we briefly discuss neutrino production in the intergalactic cascade scenario.

\section{Numerical approach}\label{sec:code}
The interplay of the physical processes discussed in the previous section governs the evolution of the particle energy distributions within the blob, and can be described by a set of time-dependent coupled integrodifferential equations.  The equation for the distribution of particle species $i$ (protons, pairs, photons, neutrons, and neutrinos) can be written in the following compact form
\eqb 
\frac{\partial n^\prime_{i}(x^\prime,\tau^\prime)}{\partial \tau^\prime} & + &  \frac{n^\prime_{i}(x^\prime,\tau^\prime)}{\tau^\prime_{i,esc}(x^\prime)} + \sum_j\mathcal{L}^{j}_i(x^\prime,\tau^\prime)  =   \nonumber \\ 
& & \sum_{j}\mathcal{Q}^j_i(x^\prime,\tau^\prime) + \mathcal{Q}^{inj}_i(x^\prime,\tau^\prime),
\label{eq:kinetic}
\eqe 
where $\tau^\prime$ is time (in units of $R^\prime/c$), $n^\prime_{i}$ is the differential number density (normalized to $\sth R^\prime$) of particle species $i$, $x^\prime$ is the particle dimensionless energy (in units of $m_e c^2$), $\tau^\prime_{i, esc}$ is the particle escape timescale (also in units of $R^\prime/c$), $\mathcal{L}^{j}_i$ is the operator for particle losses (sink term) due to process $j$, $\mathcal{Q}^j_i$ is the operator of particle injection (source term) due to process $j$, and $\mathcal{Q}^{inj}_i$ is the operator of a generic external injection. The coupling of the equations happens through the energy loss and injection terms for each particle species \citep[for their explicit form, see][]{DMPR12}. With this numerical scheme, energy is conserved in a self-consistent way, since all the energy gained by one particle species has to come from an equal amount of energy lost by another particle species.

To simultaneously solve the coupled kinetic equations for all particle types we use the time-dependent code described in \citet{DMPR12}. Photomeson production processes are modeled using the results of the Monte Carlo event generator {\sc sophia}~\citep{SOPHIA2000}, while the Bethe-Heitler pair production is similarly modeled with the Monte Carlo results of \citet{Protheroe1996} and \citet{mastetal05}. The only particles that are not modeled with kinetic equations are muons, pions, and kaons~\citep{DPM14,petroetal14}. Their energy losses and photon production via synchrotron radiation can be safely ignored for the main part of our study (Sections~\ref{sec:modeling} and \ref{sec:longterm}), but they are taken into account when discussing neutrino production from the blazar core in Section~\ref{sec:core}.

The numerical results presented in Sections~\ref{sec:modeling} and \ref{sec:other} are computed by  solving the system of equations (\ref{eq:kinetic}) for a constant injection rate of electrons and protons, $\mathcal{Q}_{e,p}^{inj}$,  and for a long enough time so that the system reaches a steady state. The steady-state approximation for modeling the blazar SED of the three consecutive days of the hard X-ray flare (Section~\ref{sec:modeling}) is valid, since the system reaches a steady state typically well within one day in the observer's frame\footnote{Only for one parameter set (Model~D), the steady state is reached in $\sim1.6$ days.}.
For the estimation of the long-term neutrino emission of the source (Section~\ref{sec:longterm}), we solve the system of equations (\ref{eq:kinetic}), using a time-dependent injection rate $\mathcal{Q}_{e,p}^{inj}$, which is motivated by the observed X-ray flux variability (details about the adopted prescription can be found in Section~\ref{sec:longterm}). By construction, a steady state cannot be reached in this case, and a time-dependent approach is more appropriate.

\section{SED modeling of X-ray flare}\label{sec:modeling}
First, we briefly describe the electromagnetic and neutrino observations used in the SED modeling of the X-ray flare (Section~\ref{sec:data}). We continue with a description of our methodology and model selection (Section~\ref{sec:param}), and present the SED modeling results in Section~\ref{sec:results}.

\subsection{Data}\label{sec:data}

The alert neutrino \icv\, was detected with the neural network classifier of~\citet{2019ICRC...36..937K}. The event was also seen with the IceCube offline follow-up selection~\citep{2019arXiv190905834M,pizzuto}. To infer the neutrino flux implied by the observation of one event with IceCube, \citet{Giommi2020} used the IceCube Alert effective area \citep{Blaufuss2019ICRC}. For completeness, we consider both the IceCube Alert neutrino effective area and the IceCube Point Source effective area~\citep{Aartsen:2018ywr} for our model predictions.

The multi-wavelength data used to describe the SED of \hsp~are taken from \citet{Giommi2020}. Specifically, the observations include pointed \swift-XRT~\citep{xrt2005} observations triggered by the IceCube alert between MJD 58856.3 (8 January 2020) and MJD 58900.5, and UVOT~\citep{uvot2005} observations from the same period. The first \swift \,  Target of Opportunity (ToO) observation of \hsp \, (obs-id: 00013051001) found the source to be in a flaring hard state: the X-ray flux was found to be $\sim2.5$ times higher than its average value in 2012-2013, and the X-ray spectrum was hard with photon index $\sim 1.8$ (see Table~2 in \citealt{Giommi2020}). The dataset also includes observations of \hsp \,  with the \nustar \, hard X-ray observatory~\citep{nustar2013} taken four days after the detection of \icv~(11 January 2020); this is the first time that \nustar \, has observed the source.

The peak frequency of the synchrotron spectrum on January 8, 2020 cannot be securely determined by the \swift \, data alone. Because of this uncertainty and the fact that the photon spectrum in the \swift-UVOT and XRT energy ranges on this day is very similar to the one on January 11, 2020, we treat both data sets as one for the purposes of the SED modeling (Section~\ref{sec:results}).

The dataset we use also includes \fermi-LAT Pass 8 data of \hsp \, from August 4, 2008 to January 8, 2020 analysed by \cite{Giommi2020}. These authors derived an average $\gamma$-ray energy flux  $F_{\gamma}=1.5^{+0.2}_{-0.1} \times 10^{-12}$~\ergcmsqs\  and photon index $\Gamma = 1.88 \pm 0.15$ in the 100 MeV--320 GeV energy range. Both estimates are consistent (within 1$\sigma$ uncertainties) with the values from the
\fermi-LAT Fourth Source Catalog Data Release 2 \citep[4FGL-DR2,][]{4FGL-DR2}, namely $F_{\gamma}=1.0\pm 0.3 \times 10^{-12}$~\ergcmsqs\ and $\Gamma = 1.89 \pm 0.17$ in the 100 MeV--100 GeV energy range. The long-term average $\gamma$-ray spectrum from \citet{Giommi2020} is included in all SED plots only for comparison purposes.
While searching for possible time-dependent $\gamma$-ray emission coincident with the X-ray flare, \citet{Giommi2020}  
also computed the \fermi-LAT spectrum of the source between MJD 58605.6 and 58855.6 which resulted in a detection with a significance (i.e., square root of the test statistic) of 2.9$\sigma$ and spectral index $\Gamma = 1.73 \pm 0.31$ (compatible with the long-term average index of the source which is $\Gamma = 1.88 \pm 0.15$). This timescale (250 days) was chosen as a compromise between achieving a detection and avoiding the wash out of possible time-dependent emission. The corresponding 250-day (long-term) photon flux integrated over the entire \fermi-LAT energy range is $1.09^{+0.96}_{-0.51} \times10^{-9}~\rm ph~cm^{-2}~s^{-1}$ ($0.60^{+0.27}_{-0.19} \times 10^{-9}~\rm ph~ cm^{-2}~s^{-1}$).

\subsection{Selection of model parameters}\label{sec:param}
In the one-zone leptohadronic model of blazar emission, the efficiency of neutrino production is a function of the target photon spectrum (spectral shape, peak frequency, and peak flux), the source radius $R^\prime$, and the Doppler factor $\dop$. When the co-moving low-energy synchrotron radiation is the main target for photomeson production\footnote{This is a good assumption for a BL Lac object \citep[for the nature of \hsp, see ][]{Giommi2020} or when the blazar zone lies outside the broad line region (BLR) of a blazar \citep[see e.g.,][for \txs]{Padovani:2019}.}, then the photomeson production efficiency ($\fpg$), defined as the ratio of the source light-crossing time and the proton energy loss timescale due to photomeson interactions, has a strong dependence on $\dop$ \citep[e.g.,][]{Murase:2014foa,petromast15}. 

\begin{figure*}
    \centering
    \includegraphics[width=0.49\textwidth]{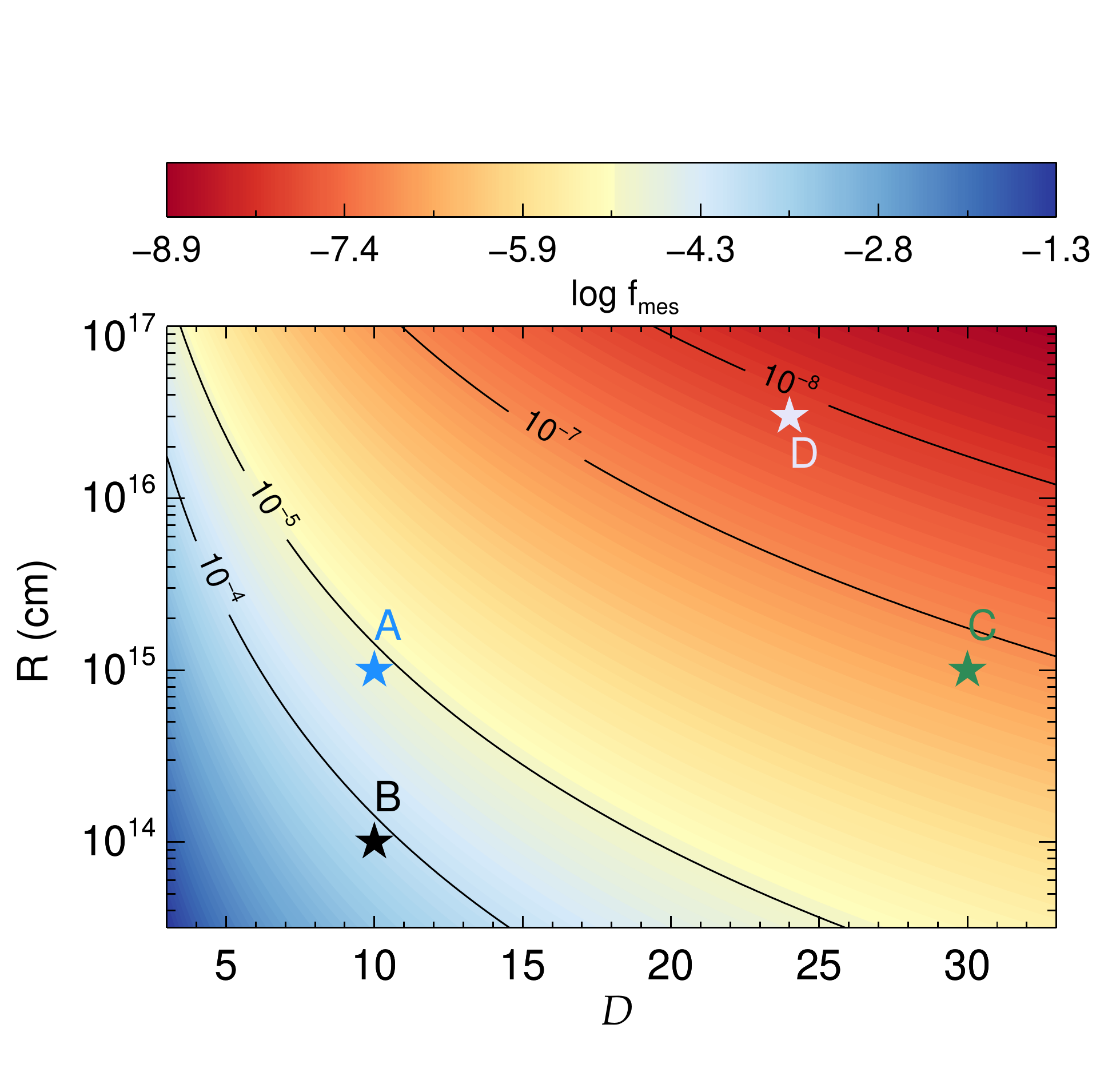}
    \includegraphics[width=0.49\textwidth]{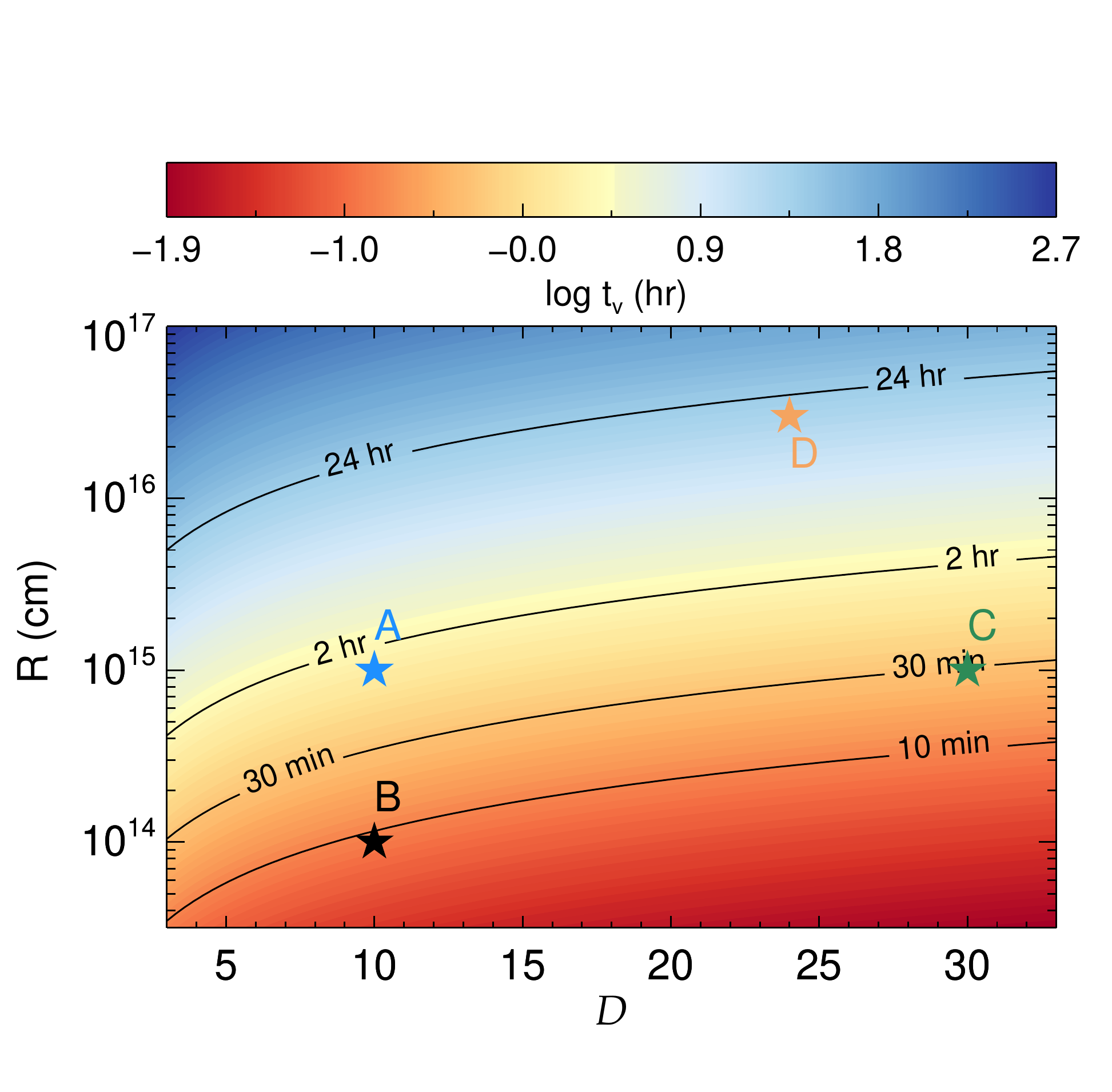}
    \caption{\textit{Left panel:} Blob radius--Doppler factor ($R^\prime-\dop$) phase space for the photomeson production efficiency, $\fpg$, of protons with the threshold Lorentz factor given by Equation~(\ref{eq:gpth}). Four indicative models, which are discussed in Section~\ref{sec:results}, are marked with stars. Contours of selected $\fpg$ values are overplotted for clarity (black lines). \textit{Right panel:} Same as in the left panel, but for the variability timescale in the observer's frame.}
    \label{fig:heatmap}
\end{figure*}
To illustrate this, we computed analytically $\fpg$ for the proton energy threshold for photomeson interactions with the peak synchrotron blazar photons of energy $\varepsilon^\prime_{s}=h \nu_s (1+z)/\dop \simeq 0.64 \,  \left(\nu_{s,18}/\dop_1\right)$~keV, where $Q \equiv Q_x 10^x$ in cgs units, unless  stated otherwise. The proton threshold Lorentz factor reads 
\eqb 
\gamma^\prime_{p,\rm th} \approx\frac{m_\pi c^2}{\varepsilon^{\prime}_s } \left(1+\frac{m_\pi}{2m_p}\right) \simeq 9\times 10^5 \, \dop_1 \nu^{-1}_{s,18}.
\label{eq:gpth}
\eqe 

In the analytical calculations, we use the step-function approximation for the cross section and a constant inelasticity of 0.2 \citep[e.g.,][]{2009herb.book.....D}. Inspired by the UV and X-ray observations of \hsp, the differential number density of the low-energy blazar photons is approximated by a broken power law with photon indices 1.7 and 2.1 below and above the break, respectively. Figure~\ref{fig:heatmap} (left panel) displays $\fpg$ (color bar) in the $R^\prime-\dop$ phase space. 

The characteristic variability timescale depends on both $R^\prime$ and $\dop$, i.e., $t_{\rm v}=R^\prime (1+z)/c \dop$, as illustrated in the right panel of Figure~\ref{fig:heatmap}. \cite{Paliya2020} report evidence for variability in the \nicer \, and \nustar \, data (taken on January 11, 2020) on timescales of $\sim 20-30$~min (at 3.5$\sigma$ and 2.2$\sigma$, respectively). \cite{Giommi2020} found no evidence for variability within individual \swift \, observations due to low photon statistics collected within the exposure time. 

Based on the above considerations, we select four pairs of $R^\prime,\dop$ values that lead to observed variability timescales ranging from $\sim 10$~minutes to $\sim1$ day, and cover a wide range of photomeson production efficiencies ($10^{-8} \lesssim \fpg\lesssim10^{-3}$). These values are marked by colored symbols in both panels, and will be used for computing benchmark leptohadronic SED models for \hsp \, (for details, see Section~\ref{sec:results}). 

For a specific choice of $R^\prime$ and $\dop$ values, one can set a lower limit on $B^\prime$, by requiring that the ratio of the synchrotron-self Compton (SSC) photon compactness to the synchrotron photon compactness\footnote{This is a dimensionless measure of the photon energy density in the source, defined as $\ell_{ph} \equiv u^\prime_{ph}\sth R^\prime /3 m_e c^2$, where $u^\prime_{ph}$ is the co-moving photon energy density.} ($\ell_{\rm ssc}/\ell_{\rm syn}$) is comparable to or lower than the so-called Compton ratio, i.e., the ratio of the observed peak $\gamma$-ray and X-ray luminosities ($L_\gamma/L_X$). This can be written as $q \equiv L_{\gamma}/L_X \gtrsim \ell_{\rm ssc}/\ell_{\rm syn} \approx \ell_{\rm syn}/\ell_B$, where $\ell_{\rm syn/\rm ssc} = \sth L_{X/\gamma} / 4 \pi R^\prime m_e c^3 \dop^4$ and $\ell_B \equiv \sth R^\prime B^{\prime 2}/8 \pi m_e c^2$ \citep[e.g.,][]{Sikora2009, 2012ApJ...749...63M, PPM2015}. 
By considering magnetic field strengths
\eqb 
B^\prime \gtrsim \sqrt{\frac{2L_X}{q R^{\prime 2} c \dop^4}} \simeq 14 \, {\rm G} \, L^{1/2}_{X, 45.5} R^{\prime -1}_{15} \dop_1^{-2} q_{-1}^{-1/2}
\label{eq:Bmin}
\eqe 
we can therefore explore models where the $\gamma$-ray emission in the \fermi-LAT band is dominated by the SSC emission of primary electrons or has a significant leptohadronic contribution  \citep{Petropoulou:2015upa, cerruti15}. In the latter case, the predicted neutrino luminosity will be higher than in the former, as demonstrated in \citet{Petropoulou:2015upa}.

After choosing values for $R^\prime, \dop$ and $B^\prime$, we can infer the properties of the primary electron distribution at injection. More specifically, we model the electron injection rate (appearing in Equation~\ref{eq:kinetic}) as a power law with a high-energy exponential cutoff
\eqb 
Q^{inj}_e = Q_{e,0}\gamma^{\prime -s_e}e^{-\gamma^\prime_e/\gamma^\prime_{e, {\rm cut}}}, \quad \gamma^\prime_e \ge  \gamma^\prime_{e, \min},
 \label{eq:Qeinj}
\eqe 
where $\gamma^\prime_{e,\min}=1$. The power-law slope $s_{e}$ can be inferred from the UV-to-X-ray spectral index $\beta$ ($F_{\varepsilon}\propto \varepsilon^{-\beta}$) as $s_e = 2\beta$ if the associated electrons are fast cooling, or $s_{e}=2\beta+1$ otherwise. \swift \, UVOT and XRT observations (see Section~\ref{sec:data}) suggest a hard power-law at injection ($s_e \lesssim 1.3$ for fast cooling electrons). In this case, the cutoff Lorentz factor, $\gamma^\prime_{e,{\rm cut}}$, is related to the observed peak synchrotron frequency $\nu_{s}$ as $\gamma^\prime_{e, {\rm cut}} \propto \sqrt{\nu_s/B^\prime \dop}$.
Finally, the co-moving injection electron luminosity $L^\prime_e \propto \int d\gamma^\prime_e \, Q_e^{inj}(\gamma^\prime_e) \gamma^\prime_e m_e c^2$ (and equivalently $Q_{e,0}$) can be inferred from the observed luminosity of the low-energy SED hump, $L_s$. For example, if electrons are fast cooling via synchrotron, then $L_e^\prime \approx L_{s}/\dop^4$.

The remaining model parameters are related to the hadronic component. In contrast to  primary electrons, the spectral shape of the relativistic proton distribution at injection cannot be inferred by the blazar SED \citep[see also][]{Keivani2018, Petropoulou2020}. We therefore assume  that the proton injection rate is described as
\eqb 
Q^{inj}_p =  Q_{p,0}
     \gamma_p^{\prime -s_p}e^{-\gamma^\prime_p/\gamma^\prime_{p,{\rm cut}}}, \quad \gamma^\prime_p \ge \gamma^\prime_{p,\min},
 \label{eq:Qpinj}     
\eqe 
where $\gamma'_{p,\min}=1$ for simplicity. 

To further reduce the number of free parameters in the model, we adopt $s_{e}=s_p$. This choice is also motivated by kinetic numerical simulations of non-thermal particle acceleration, which show that it is possible to produce electron and proton power-law energy spectra with similar slopes, depending on the physical conditions, such as the total plasma magnetization $\sigma$. For example, magnetic reconnection in electron-proton plasmas with $\sigma>1$ (relativistic regime) yields non-thermal energy spectra for both electrons and protons with similar power-law slopes \citep[e.g.,][]{2016ApJ...818L...9G}, while reconnection in plasmas with $\sigma\lesssim 1$ (trans-relativistic regime) produces power-laws with $s_p \gtrsim s_e$ \citep[e.g.,][]{2018ApJ...862...80B, 2018MNRAS.473.4840W, 2019ApJ...880...37P}. Non-thermal acceleration of electrons and protons can also take place in weakly magnetized relativistic shocks (with  $\sigma<10^{-3}$), with the produced power laws having similar slopes \citep{2011ApJ...726...75S, 2013ApJ...771...54S}.

We also set $\gamma^\prime_{p,{\rm cut}}\approx 2 \gamma^\prime_{p,\rm th}$ \citep[see also][]{Petropoulou:2015upa}. The energy of neutrinos produced by protons with Lorentz factor $\gamma^\prime_{p,\rm th}$ is approximately $\varepsilon_{\nu, \rm th} \simeq 0.4~{\rm PeV} \, \dop_1^2 \nu_{s, 18}$. If the proton distribution was extending to $\gamma^\prime_{p,\rm cut} \gg \gamma^\prime_{p, \rm th}$, then the peak energy of the neutrino spectrum would be shifted to  $\varepsilon_\nu \gg 1$~PeV. Meanwhile, the average expected energy of \icv \, lies somewhere between 0.16 and 1.4 PeV, depending on the assumed neutrino energy spectrum \citep{Giommi2020}. Finally, to derive the proton injection luminosity, $L^\prime_p \propto \int d\gamma^\prime_p \, Q_p^{inj}(\gamma^\prime_p) \gamma_p^\prime m_p c^2$, we require that the combined emission of primary electrons and secondary pairs is consistent with the broadband data.

We select an initial set of parameter values based on the analytical considerations described above. We then  perform a series of numerical simulations, as described in Section~\ref{sec:code}, with parameter values lying close to this initial set, until we obtain a reasonably good description of the SED. We report those parameters values for which the model curve passes through most of the instrument-specific SED bands, while being consistent with as many upper limits as possible. This eyeball method, which is widely adopted in blazar modeling studies \citep[e.g.,][]{Tavecchio2010, Abdo2011, Boettcher:2013wxa, cerruti15, Petropoulou:2015upa}, is sufficient for making robust predictions for the source neutrino emission. 
 
\begin{figure*}
    \centering
\includegraphics[width=0.47\textwidth]{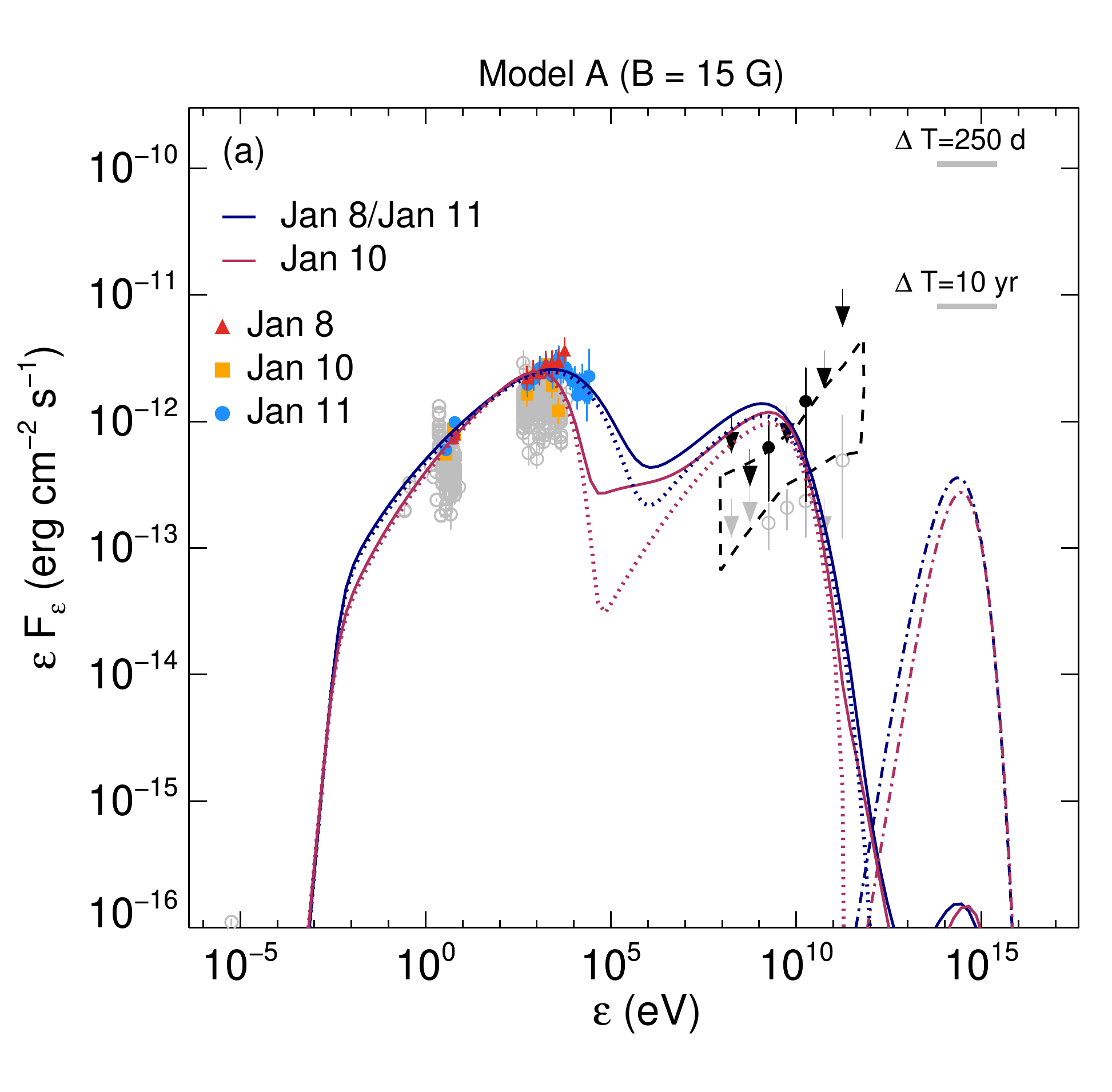}
\includegraphics[width=0.47\textwidth]{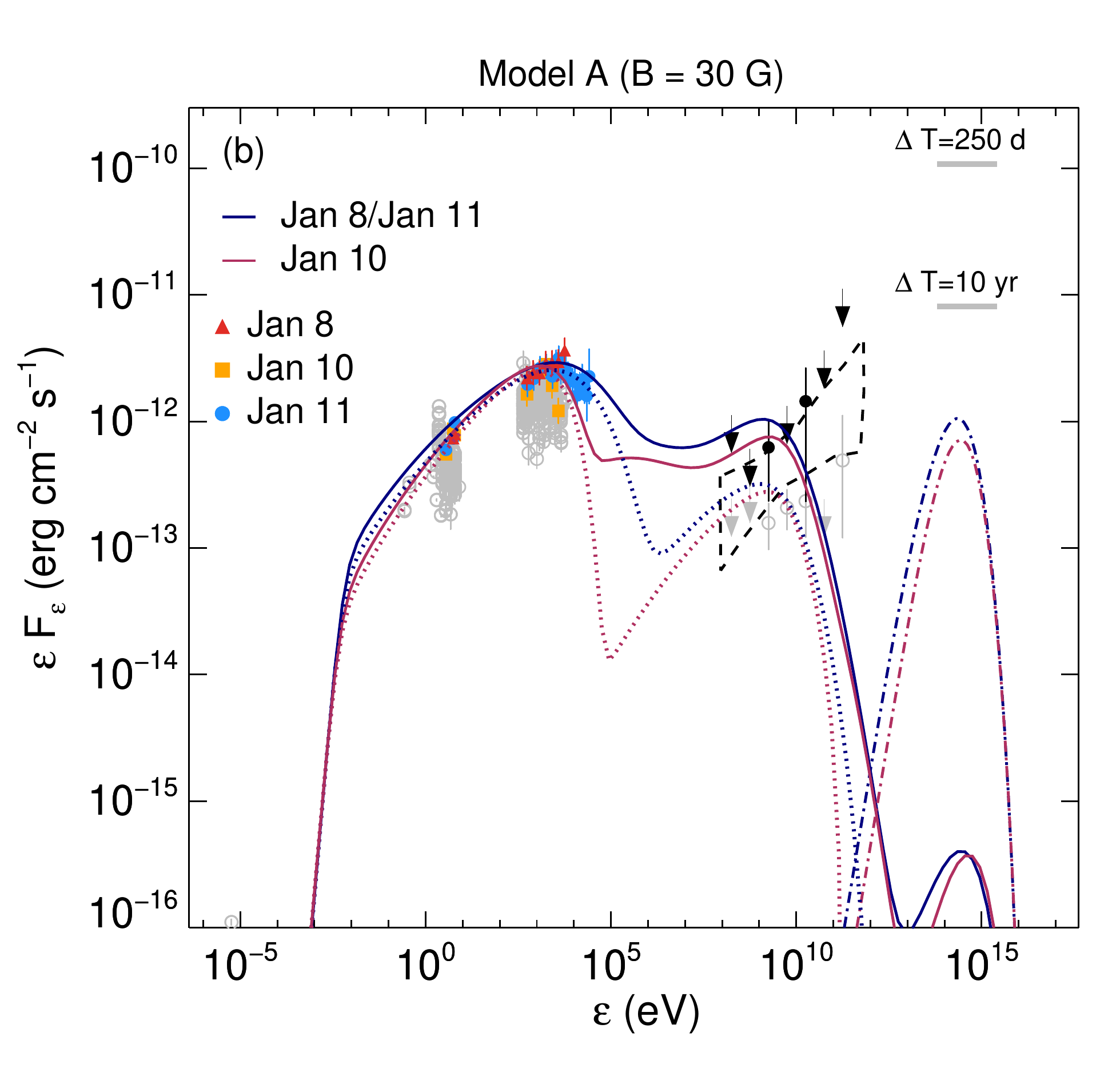}
\includegraphics[width=0.47\textwidth]{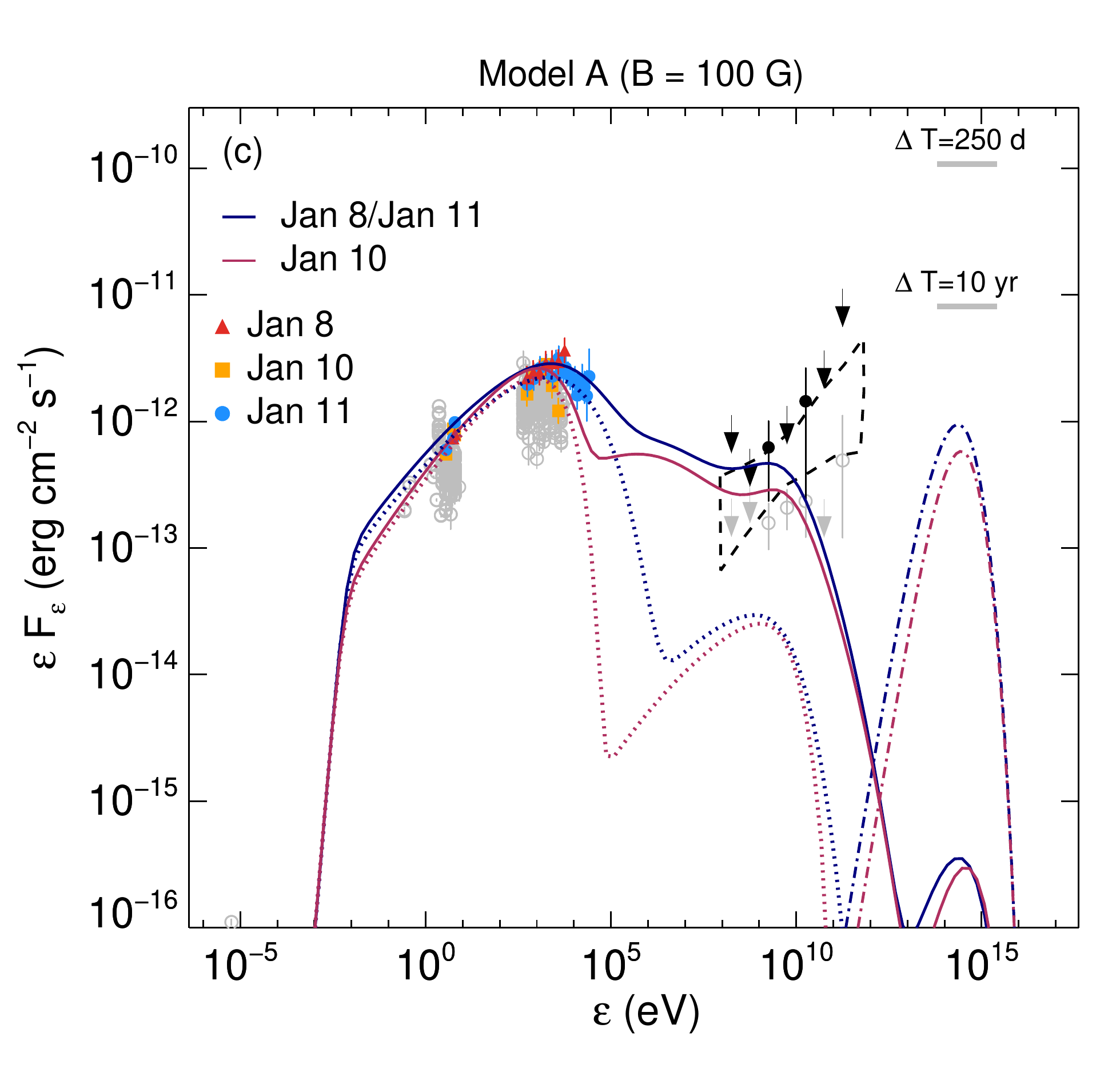}
\includegraphics[width=0.47\textwidth]{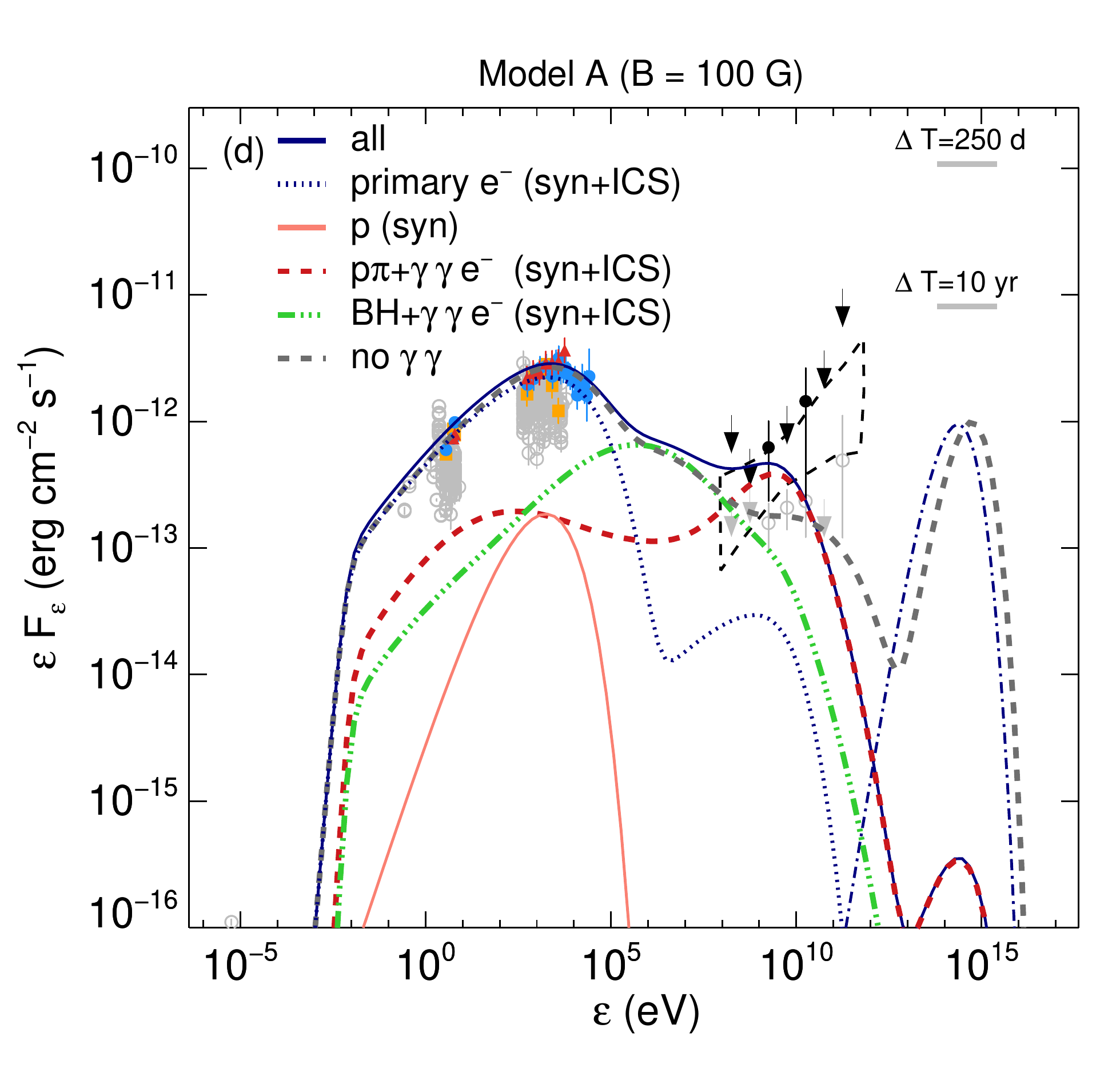}
    \caption{SEDs of \hsp \ built with data from \citet{Giommi2020}. Colored filled symbols indicate observations taken soon after the arrival of the neutrino alert (see inset legend). The inferred all-flavor neutrino flux (assuming an $\varepsilon_\nu^{-2}$ spectrum) is also marked on the plot (horizontal grey lines) for an assumed duration $\Delta T$ of neutrino emission. The black bowtie and black filled symbols show the time-integrated \fermi-LAT data over a period of 250 days prior to the neutrino alert. Archival data are overplotted with grey open symbols. In panels (a)-(c), we show the photon spectra computed in the framework of a one-zone leptohadronic model (solid lines), for three values of the magnetic field strength, as indicated on the top of each plot. The  all-flavor neutrino fluxes from the leptohadronic model are also shown in each panel (dashed-dotted lines). For comparison purposes, we also show the photon emission of primary electrons alone (dotted lines). For the parameters used, see Table~\ref{tab:param} under the column model A. Panel (d) shows the decomposition of a representative model SED into various emission components (for details, see inset legend). For clarity, we only show the spectrum and its components for January 8 and 11, 2020. In all panels, photon attenuation by the extragalactic background light (EBL) is not taken into account.}
    \label{fig:sed}
\end{figure*}
 
\subsection{Results}\label{sec:results} 
\begin{figure*}
    \centering
\includegraphics[width=0.47\textwidth]{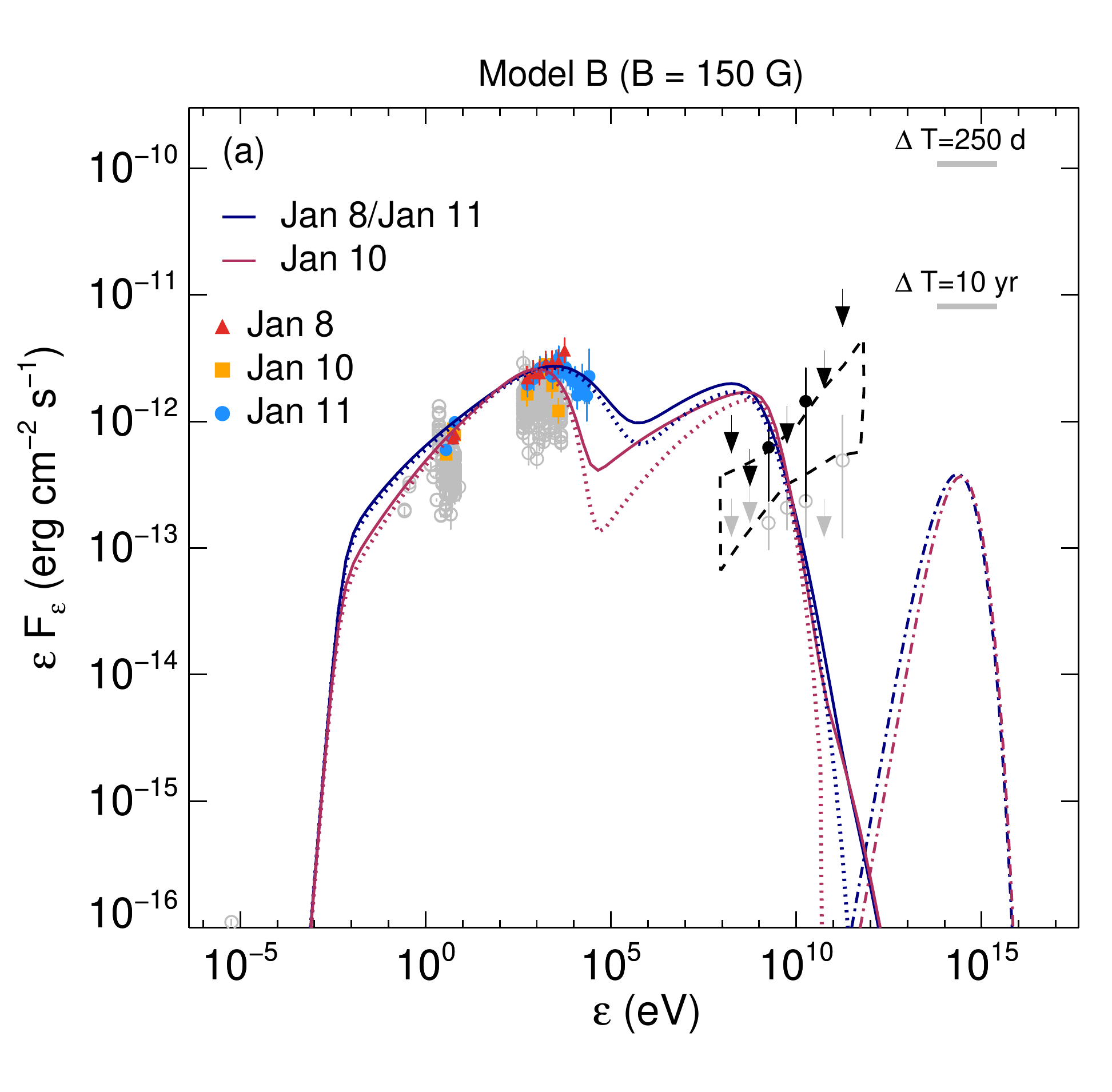}
\includegraphics[width=0.47\textwidth]{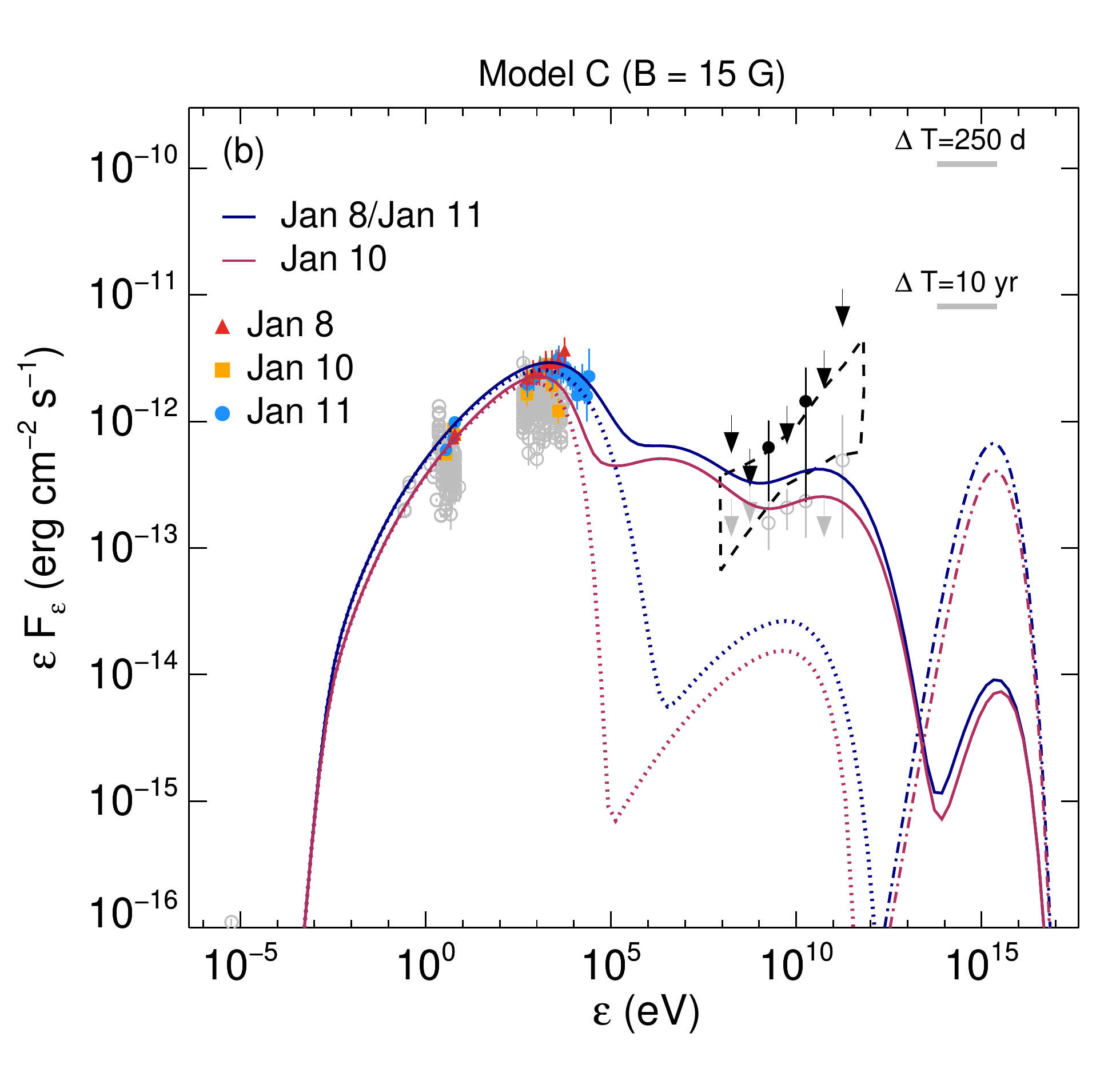}
    \caption{Same as in Figure~\ref{fig:sed}, but for Models B and C.}
    \label{fig:sed-2}
\end{figure*}

\begin{figure}
    \centering
\includegraphics[width=0.47\textwidth]{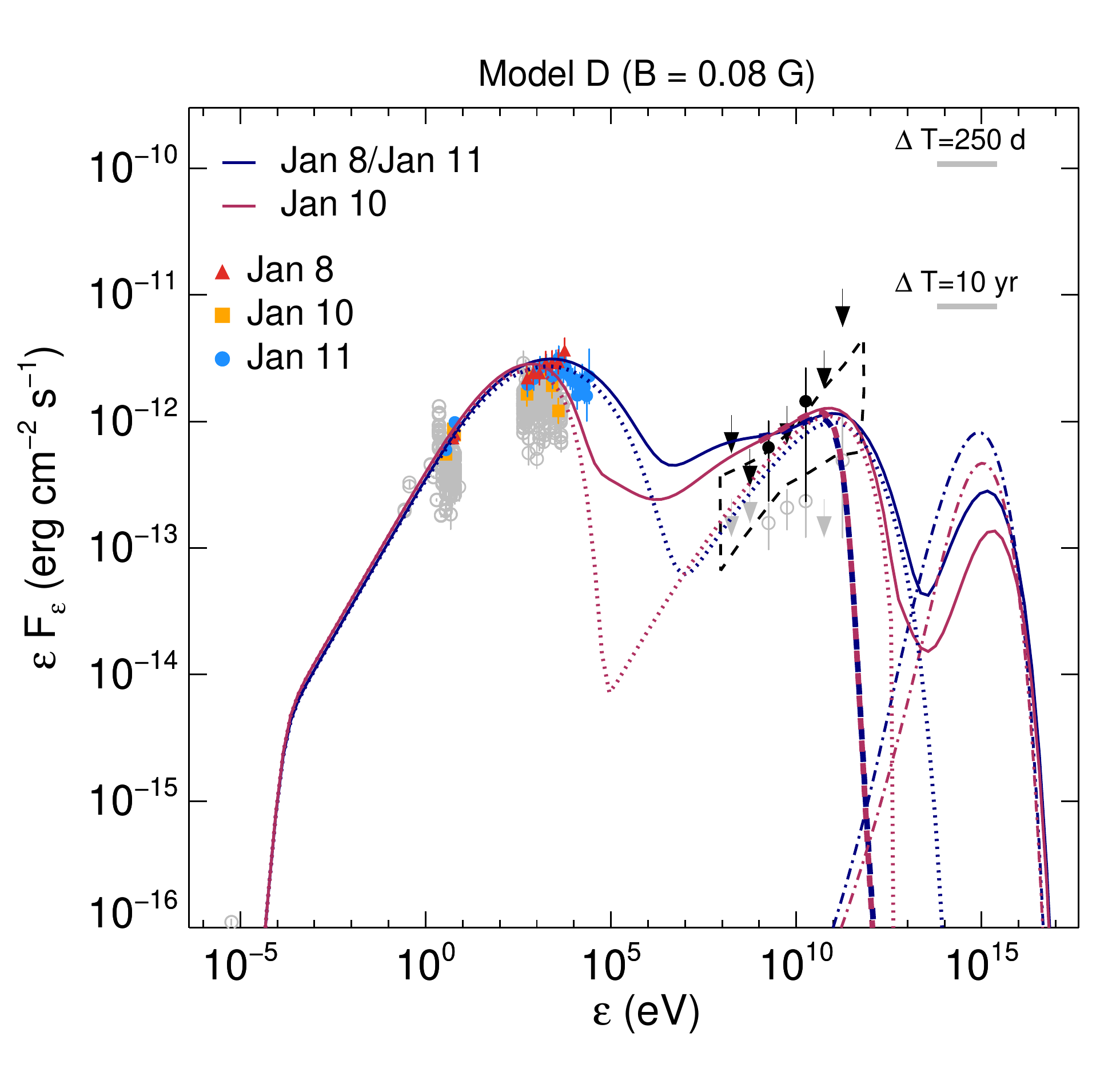}
    \caption{Same as in Figure~\ref{fig:sed}, but for Model D. For illustration purposes, we also show the EBL attenuated spectra (thick dashed lines) for the EBL model of \cite{2010ApJ...712..238F}.}
    \label{fig:sed-3}
\end{figure}

The photon and neutrino spectra computed for the epoch of the X-ray flare (January 8-11, 2020) in Models A-D are presented in Figures~\ref{fig:sed}, \ref{fig:sed-2}, and \ref{fig:sed-3}.  The input model parameters and their values are summarized in Table~\ref{tab:param}. 

Panels (a)-(c) in Figure~\ref{fig:sed} illustrate the role of the magnetic field on the predicted neutrino emission. For the selected $R^\prime$ and $\dop$, $B^\prime=15$~G (panel a) is the minimum value of the magnetic field that can yield results consistent with the observed Compton ratio (see  Equation \ref{eq:Bmin}). The $\gamma$-ray emission in this model arises mostly from the synchrotron-self Compton emission of primary electrons in the source (dotted lines). As a result, any emission originating (directly or indirectly) from photohadronic interactions can only have a minor contribution to the $\gamma$-ray emission. By increasing the magnetic field strength of the emission region (panels b and c), the SSC emission is being suppressed, thus allowing for a larger photohadronic contribution  to the overall SED. This translates to a higher proton injection luminosity (see Table~\ref{tab:param}), and is reflected in the neutrino spectrum, whose flux is also increasing (compare panels a to c). 
Additionally, the $\gamma$-ray spectrum becomes softer in the \fermi-LAT energy, with the one computed for $B^\prime=100$~G (panel c) being barely consistent with the time-integrated (yet non-contemporaneous) \fermi\, spectrum (black bowtie and symbols).

\begin{deluxetable*}{l cccccc} 
\centering
\tablecaption{Parameter values for three indicative leptohadronic models of the X-ray flare of \hsp. \label{tab:param}}
\tablewidth{0pt}
\tablehead{
\colhead{Parameter} & \multicolumn{5}{c}{Value} \\
\hline
& \multicolumn{3}{c}{Model A} & \colhead{Model B} & \colhead{Model C}   & \colhead{Model D}         
           }
\startdata
$R^\prime$ (cm) & \multicolumn{3}{c}{$10^{15}$} & $10^{14}$ & $10^{15}$ & $3\times10^{16}$\\
$\dop$ & \multicolumn{3}{c}{10} & 10 & 30 & 24 \\ 
$B^\prime$ (G) & 15 & 30 & $100$ & $150$ &  15  & $0.08$\\
$\gamma^\prime_{p,{\rm cut}}$ & \multicolumn{3}{c}{$3.2\times10^5$} & $3.2\times10^5$ & $10^6$ & $10^6$ \\ 
\hline
& \multicolumn{5}{c}{\textit{January 8 and 11}} \\
$L^\prime_e$ ($10^{42}$ erg s$^{-1}$) & 3.7 & 2.9 & 2.3 & 4.6 & $3.3\times10^{-2}$ &  $5.5\times10^{-1}$  \\ 
$\gamma^\prime_{e,{\rm cut}}$ & $10^5$ & $8\times10^4$ & $4\times10^4$ & $3\times 10^4$ &  $5\times10^4$ & $3\times10^6$\\
$s_{e}$ & \multicolumn{3}{c}{1.2} & 1.2 & 1.2 & 2  \\ 
$L^\prime_p$ ($10^{45}$ erg s$^{-1}$) & 2.7 & 5.4 & 6.8 & 0.27 & 1.7 & $5.1\times10^2$ \\ 
\hline 
& \multicolumn{5}{c}{\textit{January 10}\tablenotemark{\dag}} \\
$L^\prime_e$ ($10^{42}$ erg s$^{-1}$) &  2.3 & 1.8 & 1.5 & 2.9 & $1.8\times10^{-2}$& $5.5\times10^{-1}$\\ 
$\gamma^\prime_{e,{\rm cut}}$ & $6.3\times10^4$ & $5\times10^4$ & $2.5\times10^4$ & $2\times 10^4$ & $4\times10^4$ & $6.3\times10^5$\\
$s_{e}$ & \multicolumn{3}{c}{1} &  1 & 1.2 &  2 \\ 
$L^\prime_p$ ($10^{45}$ erg s$^{-1}$) & 2.7 & 3.4 & 4.3 & 0.27 & 1.7 &  $5.1\times10^2$ \\ 
\enddata
\tablecomments{Other parameters used in all models are: $\gamma^\prime_{e, \min}=1$, $\gamma^\prime_{p,\min}=1$, and $s_{p}=s_{e}$.}
\tablenotetext{\dag}{The electron injection rate (Equation \ref{eq:Qeinj}) is modeled with a sharp cutoff at $\gamma^\prime_{e,{\rm cut}}$ to account for the steep \swift-XRT spectrum above 1~keV.}
\end{deluxetable*}


\begin{deluxetable*}{l cccccc} 
\centering
\tablecaption{Derived physical quantities for the leptohadronic models of  \hsp \, on January 11, 2020. \label{tab:param-2}}
\tablewidth{0pt}
\tablehead{
\colhead{Parameter} & \multicolumn{5}{c}{Value} \\
\hline
& \colhead{Model A$_{(B^\prime=15 {\rm G})}$} & \colhead{Model A$_{(B^\prime=30 {\rm G})}$} & \colhead{Model A$_{(B^\prime=100 {\rm G})}$} &  \colhead{Model B} & \colhead{Model C}    & \colhead{Model D}      
           }
\startdata
$L_{\nu+\bar{\nu}}$ ($10^{45}$ erg s$^{-1}$) & 2.5 & 2.6 & 2.3 & 0.9 & 2.3 & 3.0 \\ 
$L_{\gamma}$ ($10^{45}$ erg s$^{-1}$) & 11.0 & 6.2 & 3.1 & 7.5 & 3.8 & 9.3 \\
$L_p$ ($10^{49}$ erg s$^{-1}$)& 2.7 & 5.4 & 6.8 & 0.27 & 138 & $1.7\times10^4$\\
$Y_{\nu \gamma}$ & 0.22 & 0.42 & 0.76 & 0.13 & 0.60 & 0.33\\ 
$\xi$ & $2.4\times10^3$ & $8.6\times10^3$ & $2.2\times10^4$ & $3.6\times10^2$ & $3.6\times10^5$ & $5.5\times10^7$\\ 
$P_{j}$ ($10^{47}$ erg s$^{-1}$) & 5.4 & 11.0 & 13.6 & 0.54 & 30.6 & $5.9\times10^3$\\
\enddata
\tablecomments{$L_{\nu+\bar{\nu}}$ is the all-flavor neutrino flux in the 0.1 -- 10 PeV energy range, $L_{\gamma}$ is the  $\gamma$-ray luminosity of the model in the 0.1--300 GeV energy range, $L_p=\dop^4 L^\prime_p$ is the isotropic-equivalent bolometric proton luminosity in the observer's frame,  $Y_{\nu \gamma}\equiv L_{\nu+\bar{\nu}}/L_{\gamma}$, $\xi \equiv L_p/L_{\gamma}$ is the baryon loading factor, and  $P_{j}\approx (8\pi/3) R^{\prime 2} c \Gamma^2 \left(u^{\prime}_p+u^\prime_B \right)$ is the absolute power of a two-sided jet with $\Gamma \approx \delta$ and $u^\prime_e \ll u^\prime_{B,p}$.}
\end{deluxetable*} 

As an illustrative example, we show the spectral decomposition of the model SED computed with $B^\prime=100$~G for January 11, 2020 (panel d). The effects of internal photon attenuation due to photon-photon ($\gamma \gamma$) pair production can be seen by comparing the solid blue and dashed grey lines. For the adopted source parameters, photons with energies $\gtrsim 10$~GeV (in the observer's frame) are attenuated and converted into ultra-relativistic electrons and positrons in the source. These pairs together with those produced directly by charged pion decays in the source radiate via synchrotron and Compton processes, producing a broad photon spectrum (dashed red line). In the absence of photomeson interactions, no  photons with energies $\gg10$~GeV would be produced, thus suppressing the injection of secondary pairs through $\gamma \gamma$ pair production. Thus, the combined emission of pairs from Bethe-Heitler (BH) and $\gamma \gamma$ pair production, which peaks in the MeV energy range (triple dot-dashed green line), is dominated by the former process. The proton synchrotron radiation, which peaks at $\sim1$~keV, makes a negligible contribution to the X-ray flux (solid pink line). Although the relative fluxes of the various spectral components change between different models, the general features shown in panel (d) are retained.

Models B and C, whose results are presented in  Figure~\ref{fig:sed-2},  are characterized by very different photomeson production efficiencies (see Figure~\ref{fig:heatmap}). Model B describes a very compact source with high photon densities, whereas Model C refers to a more extended source with much lower photon densities due to the adopted high Doppler factor.  The magnetic field strength used in Model B is the minimum value set by Equation (\ref{eq:Bmin}), and therefore bears similarities with Model A with $B^\prime=15$~G (panel a in Figure~\ref{fig:sed}). Because of the high photomeson production efficiency, the proton luminosity is the lowest of all models (see Table~\ref{tab:param}). Higher proton luminosities (and neutrino fluxes) would be possible in Model B for even stronger magnetic fields, as demonstrated in Figure~\ref{fig:sed} for Model A. Because of the very low photomeson production efficiency of Model C ($\fpg\sim 10^{-7}$), the optical depth for $\gamma \gamma$ pair production is accordingly low. This is also reflected in the $\gamma$-ray spectrum which for this model extends to $\sim100$~GeV. Notice also that the residual $\gamma$-ray bump from the $\pi^0$-decay is much brighter than in other models (see Figure~\ref{fig:sed}). 

The results of the fourth model we considered are presented in Figure~\ref{fig:sed-3}.  Model D is characterized by $\sim$day-long variability timescale and has the lowest photomeson production efficiency of all models (see Figure~\ref{fig:heatmap}). Because of the larger radius and higher Doppler factor, the magnetic field strength adopted here is $8\times10^{-2}$~G, i.e., close to the minimum value set by Equation (\ref{eq:Bmin}). Similarly to Model A (with $B^\prime=15$~G) and Model B (see panel a in Figures~\ref{fig:sed} and \ref{fig:sed-2}), the $\gamma$-rays are dominated by the SSC emission of primary electrons. Because of the adopted source parameters (e.g., weaker magnetic field and higher electron cutoff Lorentz factor), the shape of the SSC spectrum agrees better with that of the time-integrated \fermi \, spectrum. The combined $\gamma$-ray emission (from primary electrons and secondaries) extends to $\sim$TeV energies because of the lower $\gamma \gamma$ opacity of the emitting region. Nevertheless, to compensate for the equivalently very low $\fpg$ value, an unrealistically high proton luminosity would be required for producing a neutrino flux similar to the other models.

A summary of several physical quantities derived by the leptohadronic models discussed here (e.g., neutrino luminosity, baryon loading, jet power and others) are summarized in Table~\ref{tab:param-2}. For a detailed discussion on these results, we refer the reader to Sections~\ref{sec:jetpower} and \ref{sec:baryon}. 

\begin{figure}
    \centering
    \includegraphics[width=0.49\textwidth]{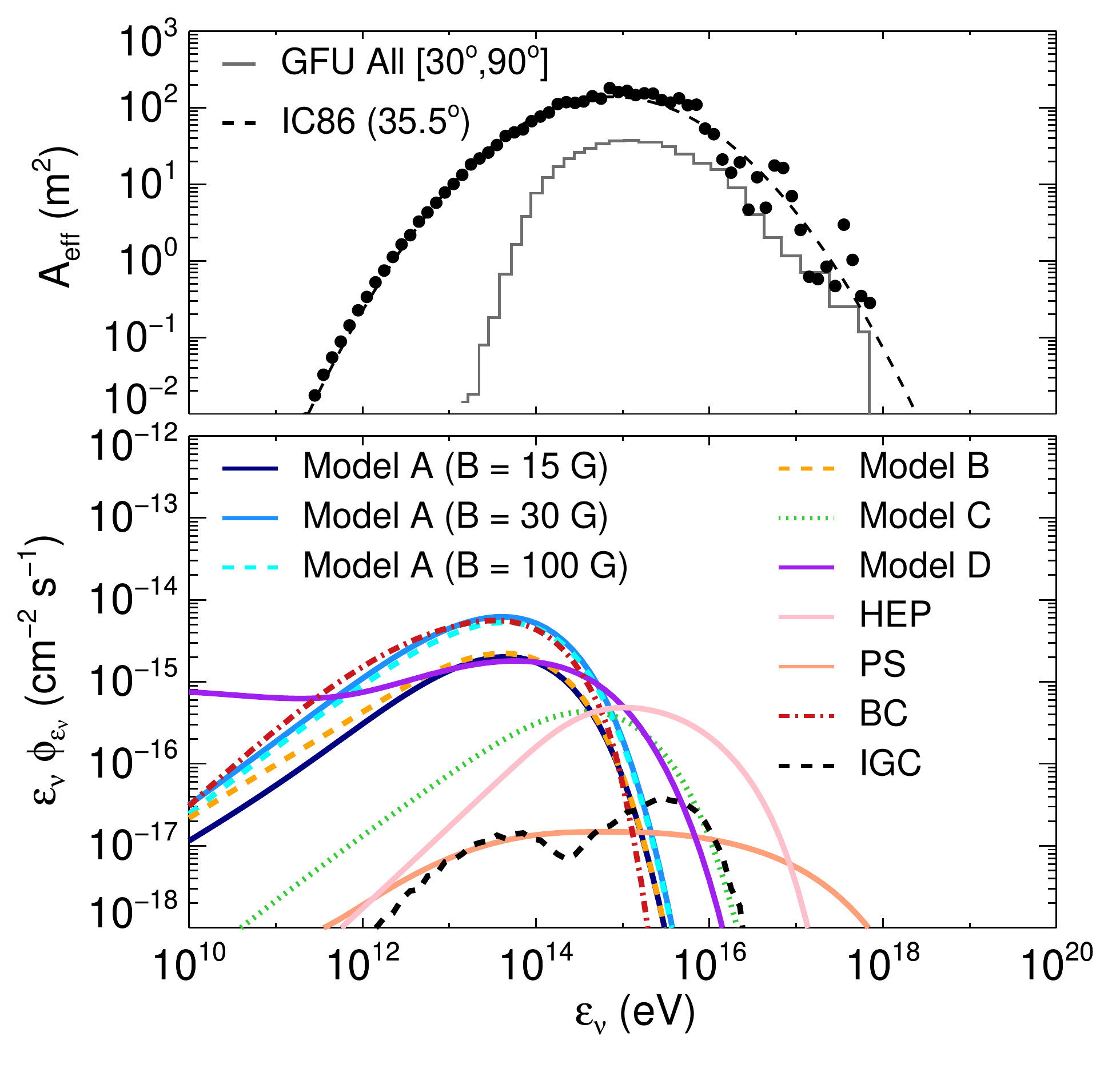}
    \caption{\textit{Top panel:} IceCube effective area of the new real-time neutrino alerts in the $[30^{\rm o}-90^{\rm o}]$ declination band (solid line) as a function of neutrino energy \citep[adopted from][]{Blaufuss2019ICRC}. The point-source effective area in the IceCube IC86 configuration at the declination of the source is also shown (filled circles show the IceCube Monte Carlo points from  \cite{Aartsen2019} and the dashed line its parameterization). \textit{Bottom panel:} All-flavor neutrino spectra predicted by the leptohadronic models of \hsp \, (for details, see inset legend). The predictions of  alternative scenarios discussed in Section~\ref{sec:other} are also shown (for ``BC'' see Section~\ref{sec:core}, for ``HEP'' see Section~\ref{sec:hep}, for ``PS'' see Section~\ref{sec:psyn}, and for ``IGC'' see Section~\ref{sec:cascade}).}
    \label{fig:Aeff-specv}
\end{figure}

We estimate next the rate of muon neutrinos and anti-neutrinos, $\dot{\mathcal{N}}_{\nu_{\mu}+\bar{\nu}_\mu}$, from the source in the neutrino emission models explored in this section, as follows
\eqb 
\dot{\mathcal{N}}_{\nu_{\mu}+\bar{\nu}_\mu} =\frac{1}{3} 
\int_{\varepsilon_{\nu, \min}}^{\varepsilon_{\nu, \max}} {\rm d} \varepsilon_{\nu} \,  A_{\rm eff}(\varepsilon_{\nu},\delta) \phi_{\varepsilon_{\nu}}.
\label{eq:Nnu}
\eqe 
Here, $\phi_{\varepsilon_{\nu}}$ is the all-flavor neutrino and anti-neutrino flux (differential in energy) of each model (computed on January 11, 2020), $\varepsilon_{\nu, \min}=100$~TeV and $\varepsilon_{\nu,\max}= \infty$ are respectively the minimum and maximum energies considered for the calculation. We also assumed vacuum neutrino mixing and use $1/3$ to convert from the all-flavor to muon neutrino flux. $A_{\rm eff}(\varepsilon_{\nu_{\mu}}, \delta)$ is the energy-dependent and declination-dependent effective area of IceCube. We have considered both the IceCube Alert neutrino effective area of~\citet{Blaufuss2019ICRC} and the IceCube Point Source effective area~\citep{Aartsen:2018ywr}\footnote{Available online at
  \url{https://icecube.wisc.edu/science/data}} in our calculations (see top panel of Figure~\ref{fig:Aeff-specv}). The fact that the IceCube Alert effective area is only available averaged in the declination range $[30^{\circ}-90^{\circ}]$ likely leads to an underestimation of the neutrino rate expected in this channel at the declination of \hsp ~by a factor of a few.  

\begin{deluxetable}{ccc}
\centering
\tablecaption{Yearly rate of muon and antimuon neutrinos expected to be detected by IceCube, and Poisson probability to detect one muon (or antimuon) neutrino with energy exceeding 100 TeV with the Alert (Point Source) search for the leptohadronic models studied  in this section.\label{tab:nu_rate}}
\tablewidth{0pt}
 \tablehead{
 \colhead{Model} & \colhead{$\dot{\mathcal{N}}_{\nu_{\mu}+\bar{\nu}_{\mu}}(> 100 ~\rm TeV)$}
 & \colhead{$\mathcal{P}|_{1\,\nu_{\mu}\,{\rm or}\,\bar{\nu}_{\mu}} (>100 ~\rm TeV)$} \\
  & ($\times10^{-4}$ yr$^{-1}$) & \\
 & Alert (Point Source) & Alert (Point Source) \\
 }
 \startdata
A$_{(B^\prime=15 {\rm G})}$ &  $17~(190)$  &  0.02~(0.2) \% \\ 
A$_{(B^\prime=30 {\rm G})}$ &  $50~(540)$  &  0.06~(0.7) \%  \\ 
A$_{(B^\prime=100 {\rm G})}$ &  $45~(490)$  &  0.05~(0.6) \%  \\ 
B &  $18~(200)$  &  0.02~(0.2) \%  \\ 
C  &  $25~(100)$  &  0.03~(0.1) \%  \\ 
D  &  $40~(210)$  &  0.05~(0.3) \%  \\ 
\enddata
\tablecomments{The rates have been computed based on the neutrino fluxes for the X-ray flare on January 11 2020, and should not be confused with the long-term predictions of Section~\ref{sec:longterm}. The Poisson probabilities are computed for a period of 44 days starting on MJD 58856.3.}
\end{deluxetable}
Table~\ref{tab:nu_rate} gives the expected number of muon and antimuon neutrinos per year in IceCube in the Alert and Point Source channels. The former is more appropriate for interpreting the recent putative association, while the latter would be appropriate for interpreting future searches by IceCube into the archival data in this direction. Although the neutrino luminosity varies only by a factor of $\sim 3$ among the models (see Table~\ref{tab:param-2}), the number of expected neutrinos varies by a factor of up to $\sim$5 because of the slightly  different spectral shapes (see bottom panel of Figure~\ref{fig:Aeff-specv}). Use of the yearly rates quoted in Table~\ref{tab:nu_rate} for computing the expected number of neutrinos in the course of $X$ years should be made with caution, since the neutrino flux associated with the X-ray flare may not be representative for the long-term neutrino emission (for details, see Section~\ref{sec:longterm}).

To summarize, we have explored four one-zone leptohadronic models for the epoch of the X-ray flare that are characterized by different source conditions, namely magnetic field strength, size, and Doppler factor. We showed that the predicted neutrino luminosity for the epoch of the X-ray flare is $L_{\nu+\bar{\nu}}=\mathcal{O}(10^{45}$~erg s$^{-1})$ (see Table~\ref{tab:param-2}), in agreement with the analytical estimates of \cite{Giommi2020}. 
The X-ray spectral changes seen above $\sim 1$~keV between January 10 and January 11, 2020 do not significantly affect the neutrino flux, as its peak value is determined by the photomeson interactions of the highest energy protons in the source with the peak synchrotron photons  in all models. Based on these results, it is unlikely that neutrino production in the jet (co-spatial with the blazar radiation zone) can yield a neutrino event, like \icv, coincident with the X-ray flare. We discuss the model implications for the long-term neutrino emission of the source in the following section.

\section{A time-dependent model for long-term neutrino emission}\label{sec:longterm}
Here, we estimate the long-term neutrino emission of \hsp \, in the context of the one-zone leptohadronic scenario. As an illustrative example, we use the parameters of Model A (with $B^\prime=30$~G) and perform time-dependent simulations of the photon and neutrino emissions by imposing temporal variations on the injection luminosities of electrons and protons. 

\begin{figure}
    \centering
    \includegraphics[width=0.47\textwidth]{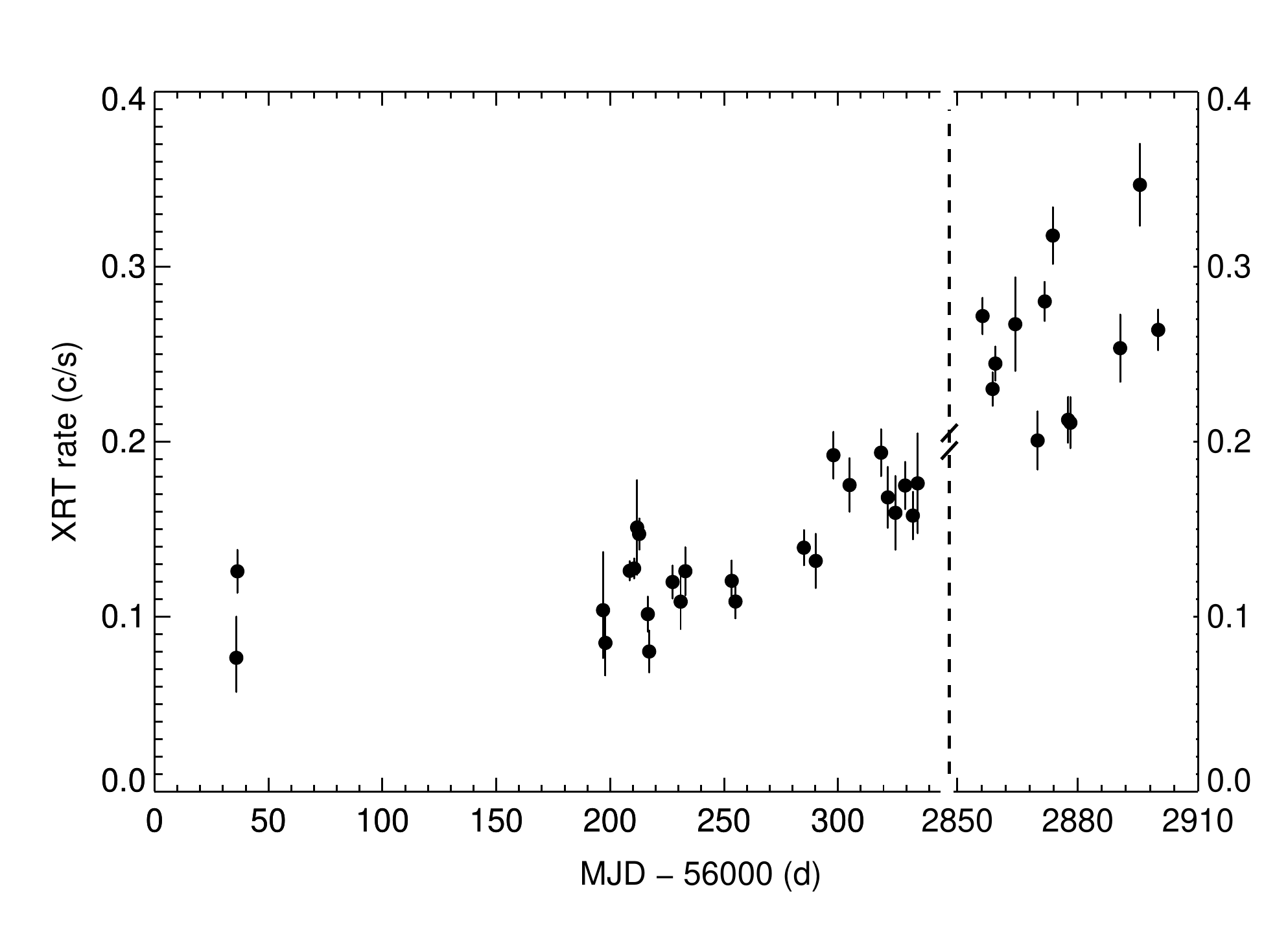}
    \caption{
    X-ray light curve of  \hsp\ (0.3--10 keV band), using all available \swift-XRT observations. 
    }
    \label{fig:xrt_lc2}
\end{figure}

X-ray photons are the main targets for photomeson interactions with protons in the source. Meanwhile, changes in the X-ray flux can be linearly mapped to changes in the electron injection rate, since the X-ray radiating electrons are fast cooling due to synchrotron radiation (this is true for all models, except for Model D). In order to determine the functional form for $L^\prime_e(t^\prime)$ and $L^\prime_p(t^\prime)$, we therefore use the \swift-XRT count rate as displayed in Figure~\ref{fig:xrt_lc2}. 
X-ray data were retrieved from the \swift \ science data centre\footnote{\url{https://swift.gsfc.nasa.gov/archive/}} and analyzed using standard procedures \citep[e.g.,][]{2019A&A...631A.116G}. Count rates were estimated from XRT images of individual observations in the 0.3--10 keV energy range.

\begin{figure}
    \centering
    \includegraphics[width=0.49\textwidth]{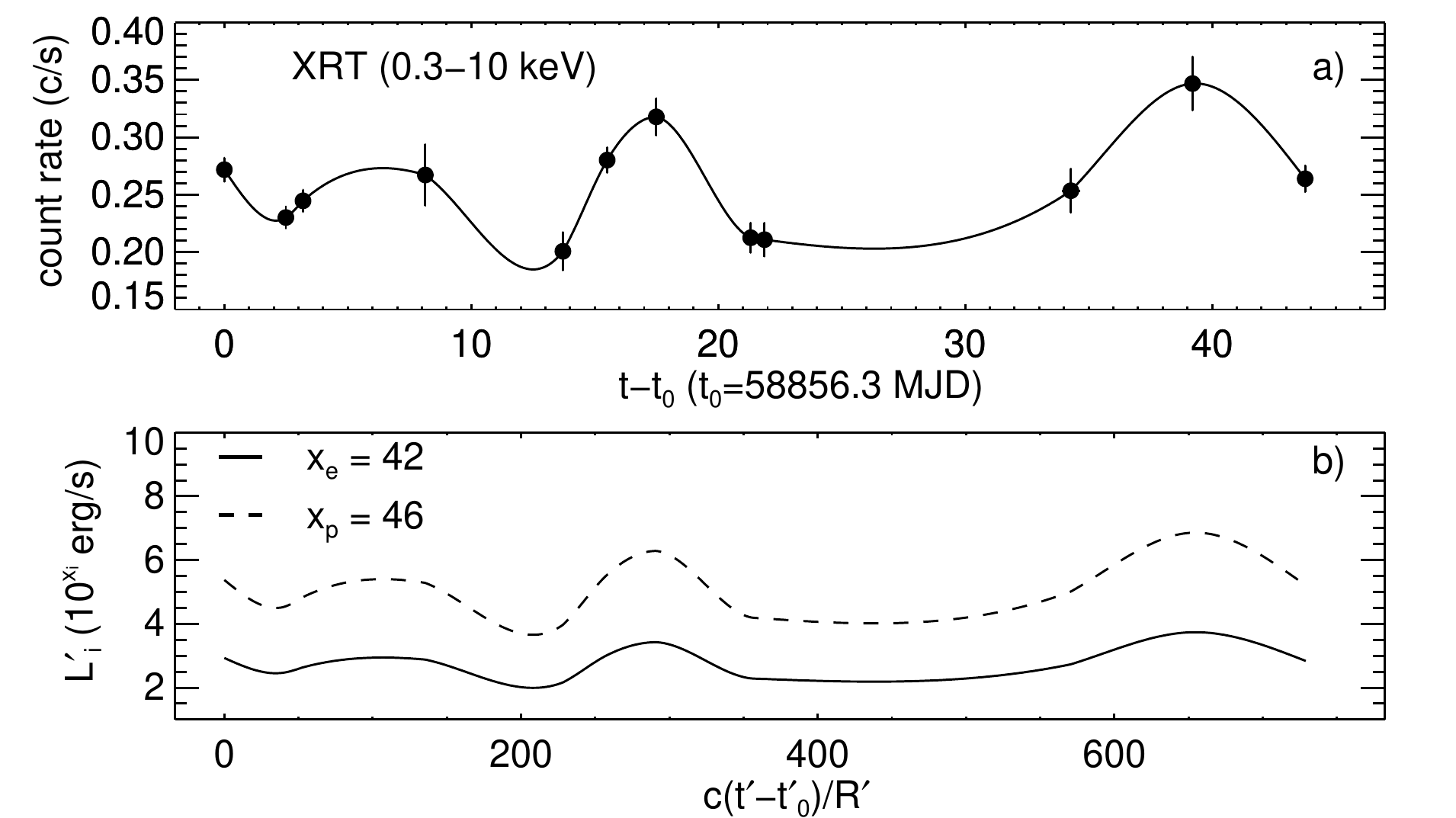}
    \caption{\swift-XRT light curve (symbols) in the 0.3--10 keV energy range (panel a) since January 8, 2020. The solid curve shows the interpolated count rate used to simulate variations of $L^\prime_e$ and $L^\prime_{p}$, normalized to $10^{42}$ and $10^{46}$~erg s$^{-1}$, respectively (panel b). All other parameters are kept fixed to their values listed in Table~\ref{tab:param} (see Model A with $B^\prime=30$~G).}
    \label{fig:xrt_lc}
\end{figure}

For simplicity, we limit our time-dependent calculations at the high-flux state after January 8, 2020 ($t_0=58856.3$~MJD), we ignore any changes in the X-ray photon index, and model both injection luminosities as
\eqb 
L_i^\prime(\tau^\prime) = \frac{CR(\tau^\prime)}{CR(\tau_0^\prime)}L_i^\prime(\tau_0^\prime)
\label{eq:lum-t}
\eqe 
where $\tau^\prime \equiv ct(1+z)/\dop R^\prime$, $\tau^\prime_0\equiv ct_0(1+z)/\dop R^\prime$, $CR(\tau^\prime)$ is the interpolated \swift-XRT count rate at co-moving time $\tau^\prime$, and $L_i^\prime(\tau_0^\prime)$ is the co-moving injection luminosity of particle species $i=e,p$ on January 8, 2020 (the values are reported in Table~\ref{tab:param} under the column for Model A with $B^\prime=30$~G). The interpolated \swift-XRT count rate curve and the variable injection luminosities of electrons and protons are shown in Figure~\ref{fig:xrt_lc}.

\begin{figure}
    \centering
    \includegraphics[width=0.45\textwidth]{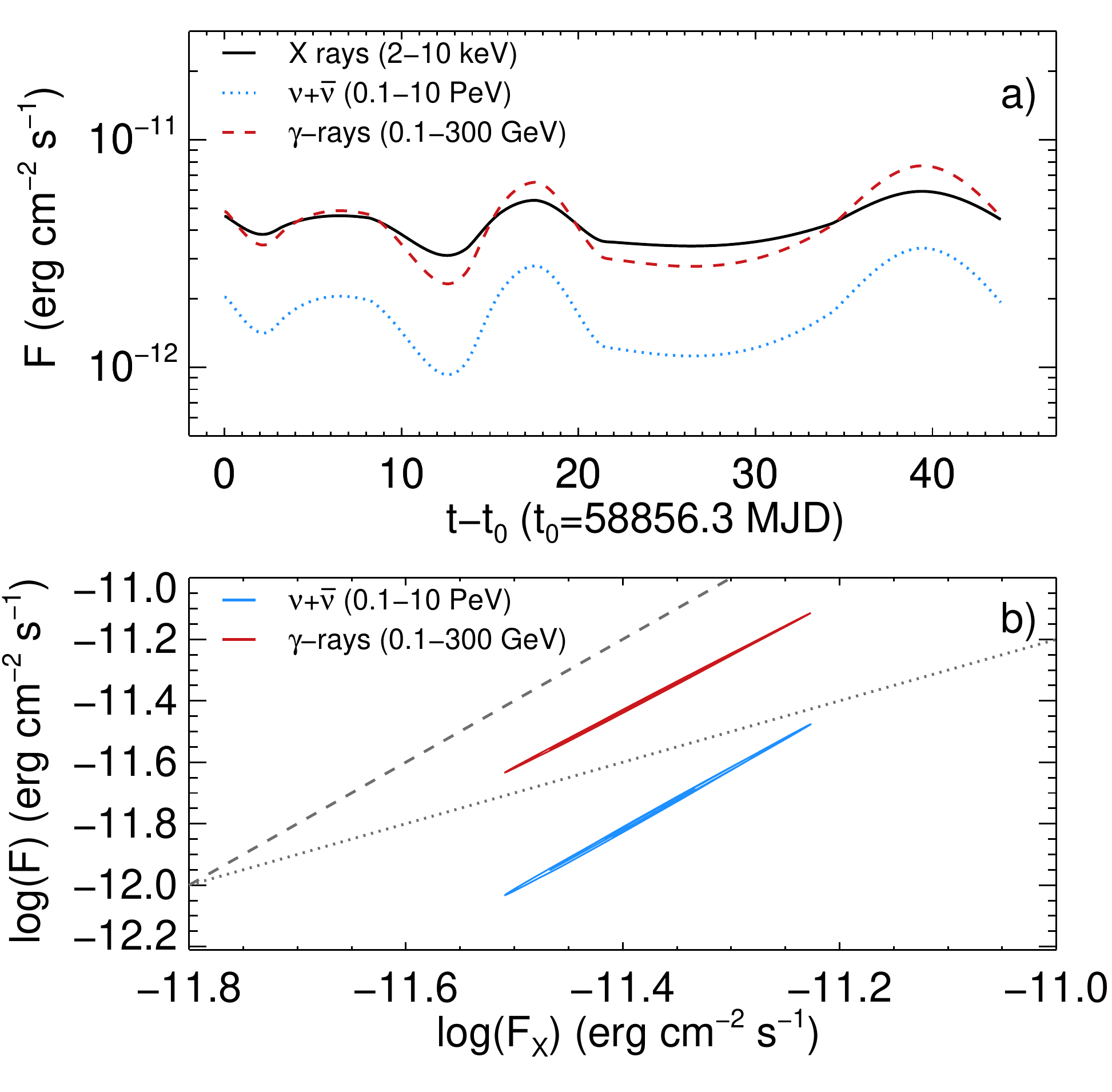}
    \caption{Results of a time-dependent model with electron and proton luminosities varying with time according to Equation (\ref{eq:lum-t}). \textit{Panel a:} Simulated 2--10 keV X-ray flux  (solid black line), 0.1--300 GeV $\gamma$-ray flux (dashed red line), and 0.1--10~PeV all-flavor neutrino flux (dotted blue line) as a function of time. \textit{Panel b:}  All-flavor neutrino  flux and $\gamma$-ray flux versus X-ray flux (in logarithmic units). Dashed and dotted lines with slopes of 2 and 1, respectively, are plotted to guide the eye.}
    \label{fig:sed-time}
\end{figure}

Using $L_i^\prime(\tau^\prime)$ as an input to the code, we simulate the time-dependent photon and neutrino emissions after January 8, 2020 for a period of $\sim44$~days ($\sim 800 R^\prime/c$) in the observer's frame (in the blob co-moving frame). 
The model-predicted X-ray flux (in the 2--10 keV energy range), the $\gamma$-ray flux (in the 0.1--300 GeV energy range), and the all-flavor neutrino flux (in the 0.1 -- 10 PeV energy range) are displayed in panel (a) of Figure~\ref{fig:sed-time}.
We find that both the $\gamma$-ray and all-flavor neutrino fluxes scale almost quadratically with the X-ray flux, as shown in panel (b) of the same figure. The quadratic dependence of $L_{\nu+\bar{\nu}}$ on $F_X$ can be understood as follows: $L_{\nu+\bar{\nu}}\propto L_p n^\prime_t \propto L_p L_e \propto L_e^2 \propto F_X^2$, where $n^\prime_t \propto L_t / \dop^4 R^{\prime 2} \varepsilon^\prime_t$ is the number density of target photons for photomeson production with energy $\varepsilon^\prime_t\approx \varepsilon^\prime_s$, and is directly proportional to $L_e^\prime$. Similarly to $L_{\nu+\bar{\nu}}$, the luminosity of other secondary particles from photomeson interactions, namely pairs from the decay of charged pions and very high-energy (VHE) $\gamma$-rays from the decay of neutral pions, will also scale as $F_X^2$. In the leptohadronic models presented in the previous section and here, the GeV flux is mostly produced by synchrotron radiation of secondary electrons and positrons that are produced via photomeson interactions both directly from the decay of charged pions and indirectly from the attenuation of VHE $\gamma$-rays from neutral-pion decay (see also panel d in Figure~\ref{fig:sed}). Thus, our numerical findings presented in Figure~\ref{fig:sed-time} (bottom panel) confirm our analytical expectations. Both scaling relations can be extrapolated to the early time XRT light curve ($t<t_0$) as well, even though this is not explicitly shown here. 

The scaling relations between $L_X$, $L_{\gamma}$, and $L_{\nu+\bar{\nu}}$ agree with the results of \cite{MPD2013}, who studied flux-flux correlations in the context of benchmark leptohadronic models  for the TeV blazar Mrk~421. More complex scaling relations can be obtained if there are spectral changes in the X-ray energy band and/or the proton injection luminosity is unrelated to that of primary electrons \citep[see also][]{MPD2013}. Interestingly, the neutrino luminosity is expected to be constant in time, if $L^\prime_p \propto L_X^{-1}$ and $L_e^\prime \propto L_X$, for all other parameters fixed.

\subsection{Average $\gamma$-ray emission}
We next estimate the long-term $\gamma$-ray flux of the time-dependent model by averaging over a period of 10 years, starting from the approximate start of IceCube operations $t_i=54557$~MJD (April 1, 2008) till $t_f=58900.5$~MJD. 
For epochs without XRT data (i.e., prior to MJD~56035.9 and MJD~56335.0--58856.3) we assumed a constant count rate equal to the mean XRT count rate from  MJD~56035.9 to MJD~56298.0 (0.121~c s$^{-1}$), when the source appeared to be in constant X-ray flux state within the uncertainties (see Figure~\ref{fig:xrt_lc2}).

The average $\gamma$-ray flux of the model can be written as 
\eqb 
\langle F_{\gamma} \rangle = \frac{\int_{t_i}^{t_f} dt\,  F_{\gamma}(t)}{t_f-t_i},
\eqe
where
\eqb 
F_{\gamma}(t) =  \left(\frac{CR(t)}{CR(t_0)}\right)^2F_{\gamma}(t_0),
\label{eq:scale-gamma}
\eqe 
with $F_{\gamma}(t_0)\simeq4.8 \times10^{-12}$~erg cm$^{-2}$ s$^{-1}$ in the $0.1-300$~GeV  energy range. This is essentially equal to the integrated flux in the $0.1-10$ GeV,  which reads $4.2\times10^{-12}$~erg cm$^{-2}$ s$^{-1}$, due to the spectral cutoff of the model. 
We find $\langle F_{\gamma} \rangle\simeq 0.9\times 10^{-12}$ ($1.0\times10^{-12}$)~erg~cm$^{-2}$ s$^{-1}$ in the $0.1-10$~GeV ($0.1-300$~GeV) energy range.
The time-integrated  \fermi \, flux (up to the time of  the neutrino alert) in the $0.1-10$ GeV ($0.1-300$~GeV) energy range is $0.7^{+0.2}_{-0.1} \, (1.5^{+0.2}_{-0.1})\times10^{-12}$~erg cm$^{-2}$ s$^{-1}$  \citep{Giommi2020}. The yearly binned 4FGL- DR2 light curve of \hsp, which contains \fermi-LAT observations from August 4, 2008 to August 2, 2018,\footnote{Available at \url{https://heasarc.gsfc.nasa.gov/W3Browse/fermi/fermilpsc.html}}~shows no significant variations during the entire period since the start of \fermi \, operations. Thus, even though the \fermi-LAT average flux quoted above is not simultaneous with the \swift-XRT observation period, it is a reasonable description of the average $\gamma$-ray flux of the source since $t_i$. Our long-term model predictions are therefore consistent with the average observed $\gamma$-ray flux. Had we adopted a higher count rate for epochs without XRT data, the average $\gamma$-ray flux of the model would be in tension with the 4FGL value.

\subsection{Cumulative neutrino number}
To estimate the cumulative number of neutrinos expected from the source in this illustrative example, we model the all-flavor (differential in energy) neutrino (and anti-neutrino) flux as
\eqb 
\phi_{\varepsilon_{\nu}}(t) =  \left(\frac{CR(t)}{CR(t_0)}\right)^2\phi_{\varepsilon_{\nu}}(t_0),
\label{eq:scale}
\eqe 
where $CR$ is the \swift-XRT count rate in the 0.3--10~keV energy range and $t_0=58856.3$~MJD. Here, we consider all available XRT data (obtained in photon count mode) from MJD~$56035.9$ to MJD~$58900.5$ (see Figure~\ref{fig:xrt_lc2}). For epochs without XRT data (i.e., prior to  MJD~56035.9 and  MJD~56335.0--58856.3) we assumed a constant count rate equal to 0.121~c s$^{-1}$, as explained in the previous section.

We apply Equation (\ref{eq:scale}) to all leptohadronic models discussed so far, since similar scaling relations between the X-ray and neutrino fluxes are expected. Even in Model D, where a sub-linear relation between $L_X$ and $L^\prime_e$ is expected due to slow synchrotron cooling of electrons ($L_e^\prime \propto L_X^\alpha$, $\alpha<1$), a quadratic relation between $L_{\nu+\bar{\nu}}$ and $L_X$ can be obtained by tweaking accordingly the proton injection luminosity (i.e., $L^\prime_p \propto L_e^{\prime 1/\alpha}$).

\begin{figure}
    \centering
    \includegraphics[width=0.47\textwidth]{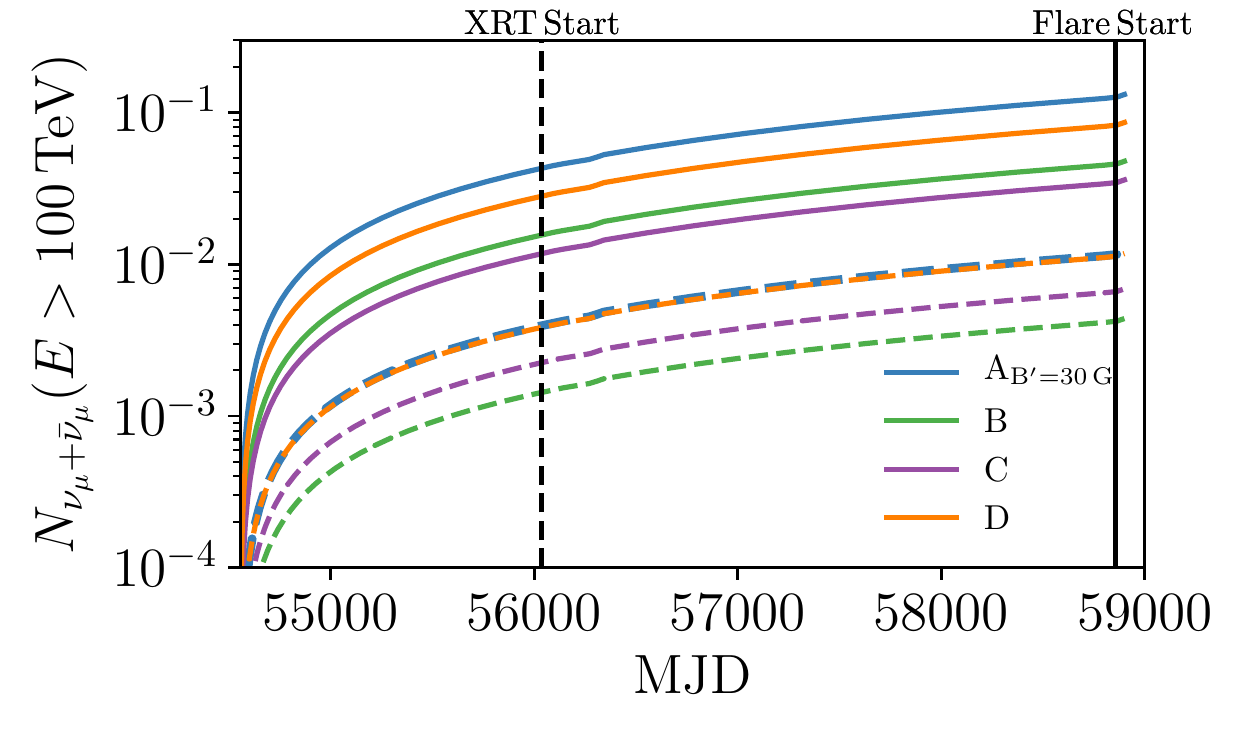}
    \caption{Cumulative number of muon neutrinos and anti-neutrinos above 100 TeV expected for IceCube with time since April 1, 2008, which roughly corresponds to the beginning of IceCube operations. The  colored solid (dashed) lines show the expected number of neutrinos using the IceCube Point Source (Alert) effective area. 
    The calculation is performed using the $\nu_\mu+\bar{\nu}_\mu$ flux estimated for Models A ($B^\prime =30\, {\rm G}$), B,C, and D, assuming that it is correlated to the \swift-XRT count rate (see Figure~\ref{fig:xrt_lc2}) according to Equation (\ref{eq:scale}). The dashed vertical line marks MJD 56035.9, the time of the first available XRT observations (denoted ``XRT Start''), and the solid vertical line marks MJD 58856.3, which corresponds to the onset of XRT observations in response to the recent X-ray flare (denoted ``Flare Start'').}
    \label{fig:neutrino_counts}
\end{figure}

Alternatively, the \fermi-LAT $\sim10$ year-long light curve of the source could be used to model the neutrino emission for the whole IceCube livetime. Given that the GeV flux variability cannot be directly tied to changes in the number density of X-ray photons, which serve as targets for photomeson interactions, or changes in the proton luminosity, one would have to make more {\sl ad hoc} assumptions about the variability patterns imposed on model parameters. As a result, the long-term neutrino predictions would be more uncertain than those made by benchmarking the model against the XRT (non-continuous) light curve.

Figure~\ref{fig:neutrino_counts} shows the expected number of neutrinos in IceCube as a function of timefrom the approximate start of IceCube operations $t_i = 54557$ MJD (April 1, 2008), 
to $t_f=58900.5$, for Models A ($B^\prime =30 \,{\rm G}$), B, C, and D, for two choices of the IceCube effective area. For clarity purposes, we do not include Model A with $B^\prime =15$~G and 100~G in the plot. In the most optimistic of the models considered, which is Model A (with $B^\prime =30 \,{\rm G}$), the expected number of neutrinos during this  $\sim$ten year-long period above 100 TeV is $N_{\nu_{\mu}+\bar{\nu}_{\mu}}(>100 ~\rm TeV)\sim0.01$ (0.1) for the IceCube Alert (Point Source) effective area.
The Poisson probability of observing one or more neutrinos when the expectation is $N_{\nu_{\mu}+\bar{\nu}_{\mu}}(> 100 ~\rm TeV) = 0.01$ is $\sim$0.01. If the neutrino detection was associated with the 44 day-long high X-ray flux state following the X-ray flare of January 8, 2020 (see Figure~\ref{fig:xrt_lc2}),  our model predicts at most $N_{\nu_{\mu}+\bar{\nu}_{\mu}}(> 100 ~\rm TeV) = 6\times10^{-4}\, (6\times10^{-3})$ with the IceCube Alert (Point Source) effective area, implying an even larger statistical fluctuation is required in order to interpret the association as physical. This finding suggests that the association of \icv~with the flare of \hsp~may be accidental.

The predicted long-term neutrino emission of \hsp~that IceCube would be expected to observe if an archival search were to be performed, is the 
flux implied by the Point Source effective area. 
 We predict $N_{\nu_{\mu}+\bar{\nu}_{\mu}}(>100 ~\rm TeV)\sim0.1$ in 10 years with our most optimistic model. For comparison, if we use the yearly rate inferred by modeling the X-ray high state of the source (see also Table~\ref{tab:nu_rate}) we predict $\sim 0.5$~$\nu_{\mu}+\bar{\nu}_{\mu}$ above 100 TeV in ten years. This is an optimistic calculation, for it assumes that the neutrino flux during the X-ray flare can be extrapolated to earlier times.
Although there is no evidence that the flare lasted that long (\hsp\, had not been observed with \swift\, prior to January 8th 2020 (MJD 58856) since December 2013 (MJD 56335)), a longer flare duration cannot be ruled out. 
Interestingly, our most optimistic long-term emission prediction is comparable to (though slightly lower than) the long-term emission of \txs~prior to 2017 calculated in~\citet{Petropoulou2020} (found to be $\sim 0.4-2$ $\nu_\mu + \bar{\nu}_\mu$ in 10 years).

\section{Other scenarios}\label{sec:other}
\begin{figure}
    \centering
    \includegraphics[width=0.44\textwidth]{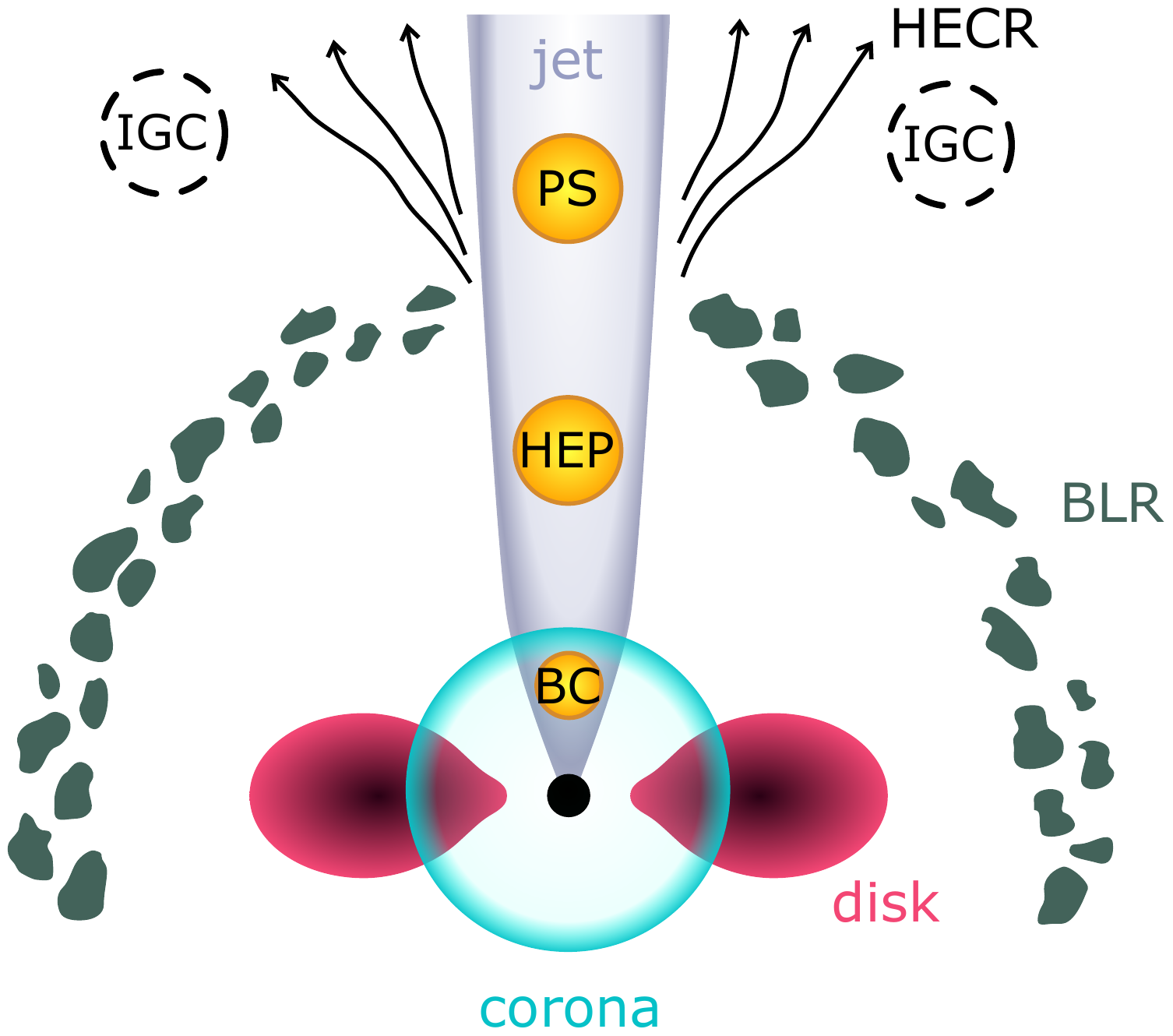}
    \caption{Schematic view of a blazar jet (grey shaded region) emerging from an accreting supermassive black hole with possible external radiation fields annotated (not in scale). Potential high-energy neutrino production sites are overplotted (blobs). The blazar-core (BC) model considers $\sim$PeV neutrino production in the inner jet (close to the accreting supermassive black hole) through interactions on coronal radiation (Section~\ref{sec:core}). The hidden external-photon (HEP) model considers $\sim$PeV neutrino production in the sub-parsec scale jet through interactions with photons from a possible weak broad line region (BLR) hidden by the jet emission (Section~\ref{sec:hep}). The proton synchrotron (PS) model assumes $\sim$EeV neutrino production through interactions of ultrahigh-energy protons on locally produced jet photons  (Section~\ref{sec:psyn}). Finally, in the intergalactic cascade (IGC) scenario, $\sim$EeV neutrinos are produced in the intergalactic medium by interactions of a high-energy cosmic-ray (HECR) beam escaping the blazar jet with the EBL photons.}
    \label{fig:sketch}
\end{figure}

In this section, we present some alternative scenarios for the neutrino emission of \hsp, where neutrino production can take place close to the supermassive black hole, or in the sub-parsec scale blazar jet, or even outside the jet (for a schematic illustration, see Figure~\ref{fig:sketch}). More specifically, we discuss a blazar-core model (Section~\ref{sec:core}), a hidden external-photon scenario (Section~\ref{sec:hep}), a proton-synchrotron model (Section~\ref{sec:psyn}), and an intergalactic cascade scenario (Section~\ref{sec:cascade}). 

\subsection{The blazar-core (BC) scenario}\label{sec:core}
We discuss a blazar-core (BC) scenario according to which the neutrino production does not take place in the blazar zone, where the bulk of the blazar's radiation originates, but occurs in the vicinity of the accreting supermassive black hole~\citep[e.g.,][]{1979ApJ...232..106E, Stecker:1991vm,2019arXiv190404226M}. GeV-TeV $\gamma$-ray emission from the core region of the active galactic nucleus (AGN) is expected to be largely attenuated, so they are often regarded as $\gamma$-ray ``hidden'' neutrino sources.   

The core itself could be thought of as part of the accretion disk and/or corona, as typically assumed in core emission scenarios for non-beamed AGN. Protons may be accelerated in the coronal region that is thought to be collisionless~\citep{2019arXiv190404226M}, and produce non-beamed high-energy neutrino and cascaded $\gamma$-ray emissions via interactions with matter and radiation from the corona. In such scenarios, the cosmic-ray proton luminosity, which is an upper bound of the expected high-energy neutrino luminosity of the source, is typically a fraction of the disk/corona luminosity. The upper limit on the bolometric disk luminosity of \hsp \, is $L_d \sim 0.5 L \lesssim 0.01 L_{\rm Edd}\sim 4\times10^{44}$~erg s$^{-1}$,  where $L$ is the (accretion-related) bolometric power derived by \cite{Giommi2020}. 
Meanwhile, the bolometric neutrino luminosity inferred by the detection of \icv, assuming a 10 year-long duration for neutrino production, is $3\times10^{46}$~erg s$^{-1}$ \citep{Giommi2020}. We can therefore conclude that a beamed neutrino source is necessary to account for this observation. In what follows, we assume that the blazar core is a relativistically moving compact region of the blazar jet, located closer to the black hole, having stronger magnetic fields and lower Doppler factors than the blazar zone.

As an illustrative example, we adopt $\dop=5$,  $B^\prime=10^4$~G,
$R^\prime=10^{14}$~cm~$\approx 2 r_g$ (where $r_g\equiv GM/c^2$), $u^\prime_p \simeq 0.2 \, u^\prime_B$, $\gamma^\prime_{p,\min}=1$, $\gamma^\prime_{p,{\rm cut}}=10^6$, and $s_p=1$. As long as $L^\prime_e \ll L^\prime_p=4\pi R^{\prime 2} cu^\prime_p/3$, the contribution of a co-accelerated electron population to the photon emission is negligible. Here, we adopted $L^\prime_e=(m_e/m_p)L^\prime_p$, $\gamma^\prime_{e,{\rm cut}}=\gamma^\prime_{p, {\rm cut}}$, and $s_e=s_p$. Contrary to the leptohadronic models for the blazar zone (Sections~\ref{sec:modeling}-\ref{sec:longterm}), we assume that the blazar-core region (being closer to the black hole) is embedded in an ambient photon field (e.g., disk corona). We model the spectrum of the {\it ad hoc} external photon field with a power-law of photon index $\Gamma=2$, extending from $\varepsilon^\prime_{\min}=10$~eV to $\varepsilon^\prime_{\max}=100$~keV, and total energy density $u^\prime_{\rm ph}=1.2\times10^4$~erg cm$^{-3}$. 
This implies that the external radiation luminosity is
$L_{\rm ph}= 4 \pi r_{\rm ph}^2 c u_{\rm ph} \gtrsim 4 \pi (R^\prime/\theta_j)^2 c u^\prime_{\rm ph}/\Gamma^2 \simeq4.5\times{10}^{43}~{\rm erg}~{\rm s}^{-1}$, for $\theta_j \approx 1/\Gamma$ and $\Gamma \approx \dop$. For simplicity, we do not include additional external radiation fields that could be related to a weak BLR, since the  $u^\prime_{\rm ext}\ll u^\prime_{ph}$ is expected (see also Section~\ref{sec:hep}).

\begin{figure}
    \centering
    \includegraphics[width=0.45\textwidth]{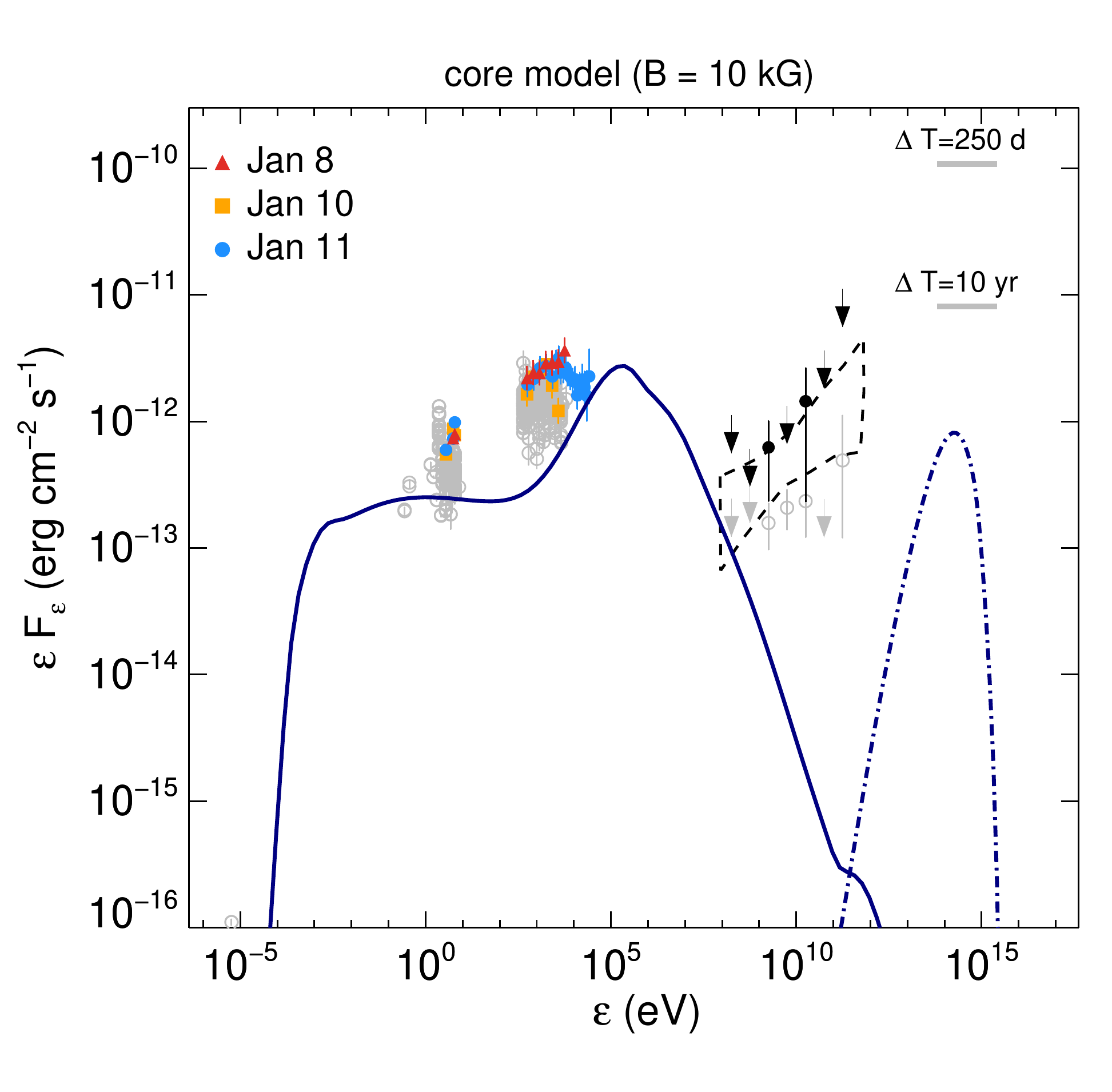}
    \caption{Photon and neutrino energy spectra (solid and dash-dotted lines, respectively) emerging from the blazar-core region. No attempt to model the blazar SED is made here, as the observed non-thermal emission is assumed to  originate from a jet region other than the blazar core.}
    \label{fig:sed-4}
\end{figure}

Under these assumptions, we compute the steady-state photon and neutrino emissions emerging from the blazar core. Because of the adopted strong magnetic field, we also take into account the synchrotron radiation of kaons, pions, and muons, as described in \cite{petroetal14}. The results of the blazar-core model are presented in Figure~\ref{fig:sed-4}. The emerging photon spectrum is mostly shaped by synchrotron radiation at low energies and $\gamma \gamma$ attenuation at higher energies ($\varepsilon \gtrsim 1$~MeV). Because the high-energy emission is re-processed to lower energies, any distinctive spectral signatures are lost~\citep[see also][]{2019arXiv190404226M}. The photon density of the hadronic-initiated cascade is comparable to that of the putative external radiation field (in the same energy range), hence the details of the latter are not important for computing the steady-state emission. 

Interestingly, the model yields a neutrino flux that is comparable to the leptohadronic models presented in Section~\ref{sec:results} (see e.g., Figure~\ref{fig:sed}). Any attempt to increase further the neutrino flux would result in even brighter electromagnetic emission that would be in tension with the low-energy tail of the \fermi \, spectrum and the hard X-ray data from \nustar \, on January 11, 2020.  In this regard, our prediction about the neutrino flux from the blazar core is the most optimistic when applied to the period of the X-ray flare. However, because the model is not designed to explain the observed SED, its predictions are not benchmarked against a specific period of interest, like the X-ray flare studied in previous sections.  Thus, persistent multimessenger emission from the blazar core is a possibility, and the model predictions can be relevant for the neutrino emission from the core on longer (year-long) timescales. In this case, however, hard X-ray data cannot be used to constrain the model due to the lack of \nustar \, observations prior to January 2020. 

\subsection{The hidden external-photon (HEP) scenario}\label{sec:hep}
In Sections~\ref{sec:modeling} and \ref{sec:longterm}, we focused on the standard single-zone models without external radiation fields. However, additional photon sources can be relevant even if they are not directly visible in the data. 

Inclusion of external photon fields has been shown to significantly enhance the efficiency of high-energy neutrino production in blazar jets \citep[e.g.,][]{Atoyan:2001ey, Dermer:2014vaa,Murase:2014foa}. Interestingly, detailed modeling of \txs \, during its multi-wavelength flare in 2017 showed that an external radiation field was necessary to explain the SED, especially when the \swift-UVOT data were taken into account~\citep{Keivani2018}.

The upper limit on the bolometric accretion luminosity of \hsp, $L/L_{\rm Edd}<0.02$, translates into an upper limit on the luminosity of a putative broad line region (BLR), as $L_{\rm BLR} \approx \xi L_d \approx \xi L/2 \lesssim 10^{-3} \xi_{-1} L_{\rm Edd}$. The upper limit on the BLR radius is estimated to be $R_{\rm BLR}\approx 10^{17} L_{d,45}^{1/2}$~cm~$\lesssim 6\times10^{16}$~cm.
Motivated by the possible presence of a weak BLR, we explore a scenario where high-energy neutrinos and $\gamma$-ray photons are produced by photohadronic interactions of relativistic protons in the jet with external photons. Lower energy radiation (from optical to X-rays) can still be produced in the same region by a co-accelerated electron population (one-zone model)~\citep[for an application to \txs, see][]{Keivani2018} or it can originate from a different part of the jet~\citep[two-zone model; for an application to \txs, see][]{Xue:2019txw}. 

Contrary to the one-zone leptohadronic models examined in the previous sections, the neutrino production site of the jet is assumed to lie within the radius $R_{\rm ext}$ of an isotropic external grey-body photon field of luminosity $L_{\rm ext}$ and effective temperature $T_{\rm ext}$. This is hidden to the observer by the non-thermal jet radiation. The photomeson production efficiency scales as $\fpg\propto \Gamma R^\prime L_{\rm ext} /R^2_{\rm ext} T_{\rm ext}$, and the neutrino luminosity will scale as $\varepsilon_\nu L_{\varepsilon_\nu}\propto \varepsilon_p L_{\varepsilon_p} \fpg$. Due to photon-photon pair production on the external photons with $\varepsilon_{\rm ext}^\prime=3 \Gamma k_B T_{\rm ext}$, there is a cutoff in the  $\gamma$-ray spectrum above an energy $\varepsilon_\gamma \approx 2 \dop (m_e c^2)^2/\varepsilon_{\rm ext}^\prime \simeq 195 (\dop_1/\Gamma_1) T_{\rm ext,4}^{-1}$~GeV
which becomes sharper with increasing values of  $L_{\rm ext}$. Protons interacting at the threshold for photomeson production with external photons of energy $\varepsilon^\prime_{\rm ext}$ (see also Equation~\ref{eq:gpth}) produce neutrinos of energy $\varepsilon_{\nu} \approx 0.05 m_p c^2 (\dop/\Gamma) (m_\pi c^2 / \varepsilon_{\rm ext}) \simeq 2.5 (\dop_1/\Gamma_1) T_{\rm ext,4}^{-1}$~PeV. If the proton distribution extends beyond $\gamma^\prime_{p, \rm th}$, then more energetic protons can interact with photons of energy $\varepsilon^\prime_{\rm ext}$ (via the multi-pion production channel), thus enhancing the neutrino flux.

\begin{figure}
    \centering
    \includegraphics[width=0.45\textwidth]{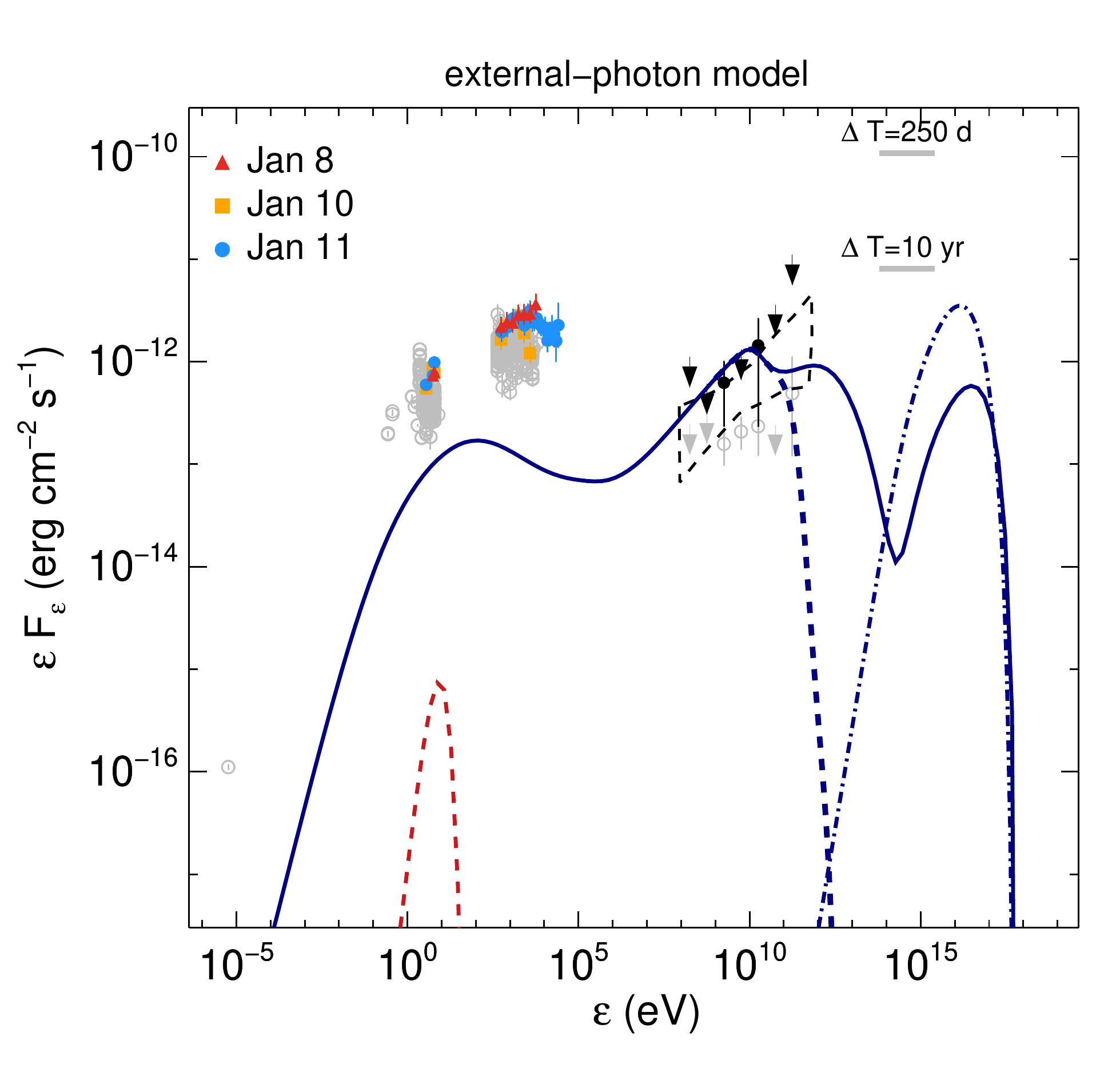}
    \caption{Photon and neutrino energy spectra (solid and dash-dotted lines, respectively) computed assuming an external radiation field (dashed red line) hidden by the jet emission. For illustration purposes, we also show the EBL attenuated $\gamma$-ray spectrum (thick dashed blue line) for the EBL model of \cite{2010ApJ...712..238F}. No attempt to model the low-energy hump of the SED is made here.}
    \label{fig:sed-5}
\end{figure}

As an illustrative example, we adopt $\dop =\Gamma=25$, $B^\prime=1$~G, $R^\prime=2\times10^{15}$~cm, $R_{\rm ext}=6\times10^{16}$~cm, $L_{\rm ext}=10^{42}$~erg s$^{-1}$, $\varepsilon_{\rm ext} = 3 k_B T_{\rm ext}\simeq10$~eV, $L^\prime_p=1.7\times10^{44}$~erg s$^{-1}$, $\gamma^\prime_{p,\min}=1$, $\gamma^\prime_{p,{\rm cut}}=3.2\times10^7\gg \gamma^\prime_{p, \rm th}\approx 6\times10^5$, and $s_p=1.5$. The jet power in relativistic protons, which is a good proxy for the total jet power in this example, is $P_j \simeq 10^{47}$~erg~s$^{-1}$. The results for the photon and neutrino emissions are depicted in Figure~\ref{fig:sed-5}.  The all-flavor peak  neutrino energy flux is $\sim 3\times10^{-12}$~\ergs, and is the highest among all considered scenarios.

 In general, the HEP scenario predicts lower neutrino fluxes by a factor of a few (depending on source parameters), if both the X-ray and $\gamma$-ray emissions originate from the same region (i.e., single-zone leptohadronic model with external photons). This can be understood as follows. Injection of primary relativistic electrons with non-negligible luminosity in the same region would contribute to the GeV flux via external Compton scattering. Thus, a lower proton injection luminosity would be required to be consistent with \fermi-LAT data, and would in turn yield lower neutrino flux. For instance, we find that the X-ray flare can be explained in a single-zone HEP scenario with the same parameters as here, and primary electrons with $L^\prime_e \approx (m_e/m_p) L^\prime_p$, but at the cost of a two times lower neutrino flux (not explicitly shown in the figure).

 Although we  tried to explain the high $\gamma$-ray state of the source in this example, the HEP scenario can also be applied to the long-term average $\gamma$-ray emission of the source. Given that in the HEP scenario the relation $\varepsilon_{\nu} F_{\varepsilon_{\nu}} \sim \varepsilon_{\gamma} F_{\varepsilon_{\gamma}}$ holds approximately, the peak neutrino flux (in $\varepsilon_{\nu} F_{\varepsilon_{\nu}}$ units)  associated with the long-term average \fermi-LAT spectrum would be lower than the one shown in Figure~\ref{fig:sed-5} accordingly.

\subsection{The proton synchrotron (PS) scenario}\label{sec:psyn}
So far, we have considered models where the high-energy emission of \hsp \, is explained by the SSC emission from primary electrons and/or the synchrotron and Compton emissions of secondary electrons and positrons produced in photohadronic interactions and photon-photon pair production. In these scenarios, the neutrino spectrum is expected to peak in the $\sim$PeV energy range \citep[see also][]{DPM14, Petropoulou:2015upa}. 

Alternatively, the high-energy blazar emission can be the result of synchrotron radiation from relativistic protons in the jet \citep{aharonian00, Muecke01}. In the proton synchrotron (PS) scenario, however, the neutrino flux is expected to peak at energies $\gtrsim 100$~PeV \citep[e.g.,][]{DPM14, Keivani2018, LP20}. This is illustrated in Figure~\ref{fig:Aeff-specv}, where we compare the neutrino spectra from the leptohadronic models with the one computed for a PS model for the X-ray flare of \hsp  \, (on January 11, 2020). Although the peak neutrino flux  (in $\varepsilon_\nu F_{\varepsilon_\nu}$ units) in the latter scenario is similar to the one computed for the leptohadronic models, 
the expected rate  of muon neutrinos in the PS model is significantly lower than in the leptohadronic models (i.e., $7\times 10^{-4}$~yr$^{-1}$ and $2\times 10^{-4}$~yr$^{-1}$ for the IceCube Point Source and Alert searches, respectively). This is a direct consequence of the much higher peak neutrino energy in the PS model (i.e., $\sim1$~EeV) and the steeply decreasing effective area of IceCube at energies $\gtrsim 1$~PeV. The PS model falls short in explaining the neutrino flux inferred by the detection of \icv,  even if the neutrino emission lasted for 10 years.  Additional high-energy neutrino emission is expected, if a fraction of the relativistic protons in the jet escape and are energetic enough to interact with EBL photons (see next subsection). 

For the PS model, we use the same parameters for the source and primary electron distribution as in Model A with $B^\prime=100$~G (see Table~\ref{tab:param}), but adopted a much higher proton cutoff energy  ($\gamma^\prime_{p, {\rm cut}}=2\times10^9$) in order to explain the $\gamma$-ray spectrum as proton synchrotron radiation. 
Meanwhile, the proton injection luminosity, which is directly related to the $\gamma$-ray emission in the proton synchrotron model, is $L^\prime_p=8.5\times10^{43}$~erg s$^{-1}$. The jet power is $P_j \simeq 2.7\times10^{46}$~erg s$^{-1}$ and is significantly lower than all leptohadronic models discussed so far (see Table \ref{tab:param-2}). Additionally, for the adopted parameters there is a rough energy equipartition between relativistic particles and magnetic fields ($u^\prime_p \sim 2 u^\prime_B$). Although the proton synchrotron scenario is strongly disfavored for the majority of blazars (particularly LBLs), it can still be energetically viable for some individual blazars (particularly, of the HBL class) as shown here \citep{cerruti15, PD16, LP20}. 

\subsection{The intergalactic cascade (IGC) scenario}\label{sec:cascade}
We finally consider the possibility that a cosmic-ray beam escapes the source, and induces an intergalactic high-energy cosmic ray (HECR) cascade. This scenario has often been proposed in connection with extreme HBLs owing to their generally hard TeV spectra and absence of TeV $\gamma$-ray variability, which is expected if the $\gamma$-rays have a secondary origin due to the deflections experienced by the parent HECRs~\citep[e.g.,][]{2010APh....33...81E,2010PhRvL.104n1102E,Essey:2010er,2012ApJ...749...63M,2013ApJ...771L..32T,2019MNRAS.483.1802T}. The indications of $\sim$year-long variability that we have seen in the \fermi-LAT spectrum of this source, if confirmed, would rule out the HECR cascade scenario as the origin of the GeV emission of \hsp. 

 \begin{figure}
    \centering
    \includegraphics[width=0.47\textwidth]{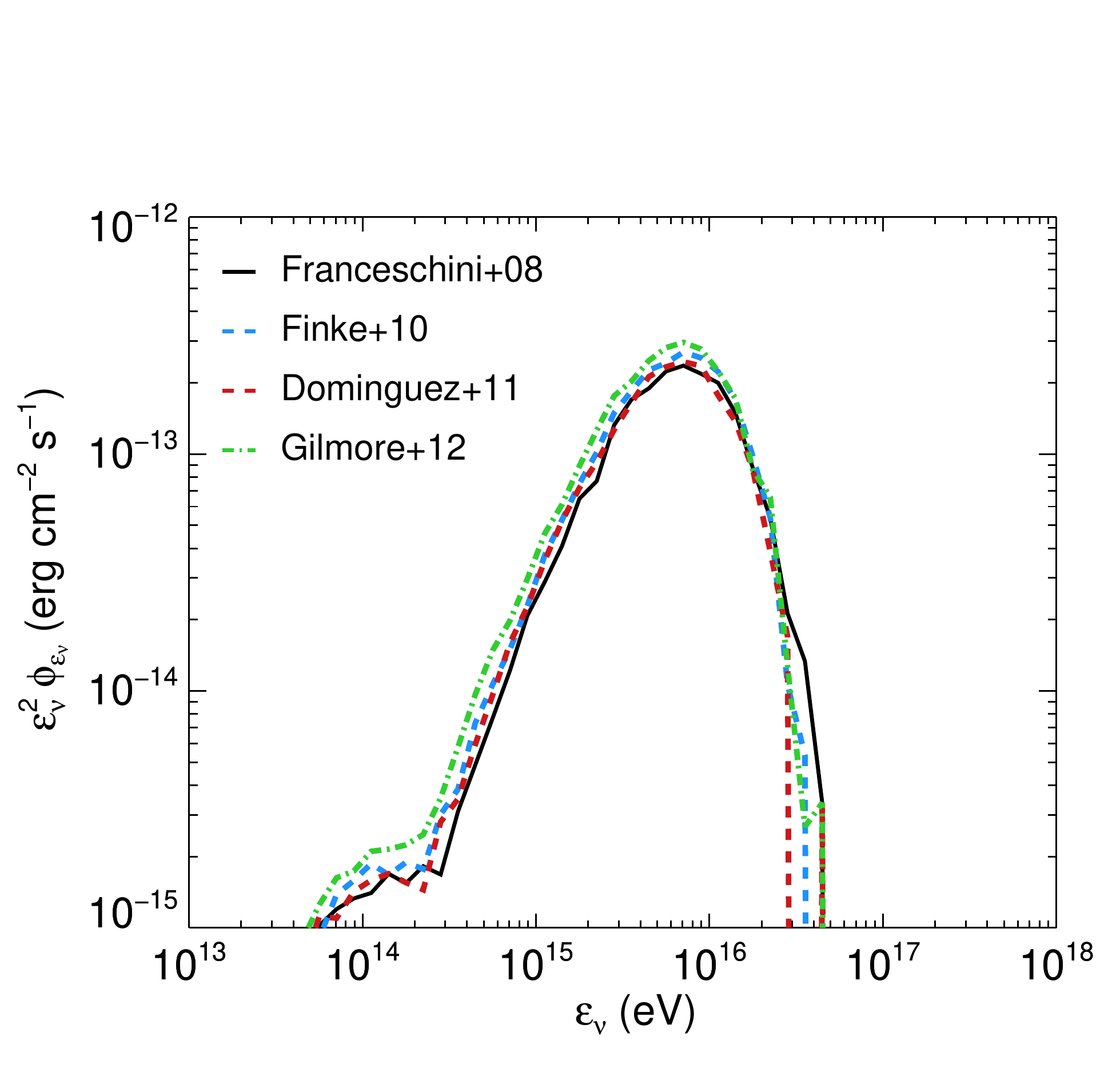}
    \caption{The expected neutrino energy flux (all flavor) in the intergalactic cascade scenario for four different different EBL models (see inset legend). The assumed isotropic-equivalent high-energy cosmic-ray luminosity is $L_{p,esc} = 3 \times 10^{49}$~erg s$^{-1}$ and the maximum proton energy is $\varepsilon_{p,\max} = 2 \times 10^{17}$~eV.}
    \label{fig:igc}
\end{figure}

We use CRPropa3~\citep{2016JCAP...05..038A} to compute the neutrino emission expected if a HECR beam escapes~\hsp, from the interactions of the cosmic rays with extragalactic background photons during their intergalactic propagation. As an illustrative example, we assume that the physical conditions in the source ($s_p$, $\dop$, $R$, $B$) are well described by Model D (the IGC scenario could in principle also apply for the parameters of Models A-C, as long as the $\gamma$-ray emission which emerges from the jet, does not already saturate the observed \fermi-LAT spectrum). We estimate the maximum proton HECR energy by equating the acceleration timescale $t^{\prime}_{\rm acc} = \eta \varepsilon^{\prime}_p/ceB^{\prime}$, where $\eta$ depends on the details of the acceleration mechanism, with the escape timescale $t^{\prime}_{\rm esc} = R^\prime/c$ (as this is shorter than the synchrotron cooling timescale). Here we adopt a fiducial value of $\eta = 100$, which yields $\varepsilon_{p,\max}=2\times10^{17}$~eV. We assume that the isotropic-equivalent escaping proton luminosity equals $L_{p,esc} = 4 \times 10^{49}$~\ergs, which is consistent with the much higher proton luminosity inside the jet of Model D (see Table~\ref{tab:param}). This corresponds to absolute, beaming corrected, proton luminosity $\mathcal{L}_p = 4 \times 10^{46}~\mathrm{erg~s^{-1}}(L_{p}/4\times 10^{49}~\mathrm{erg~s^{-1}})(\Gamma/24)^{-2}$, comparable to the Eddington luminosity of the $3 \times 10^{8} M_{\odot}$ black hole~\citep{2020MNRAS.495L.108P}.  

We do not include the effect of the intergalactic field, which would deflect some of the HECRs out of the line of sight, and reduce the expected neutrino signal. We investigated the effect of the choice of EBL model, and find the expected neutrino flux to be very robust to this model uncertainty. The predicted neutrino spectra emerging from the IGC scenario are shown in Figure~\ref{fig:igc}. For all four EBL models explored~\citep{2011MNRAS.410.2556D,2010ApJ...712..238F,2008A&A...487..837F,2012MNRAS.422.3189G}, the neutrino flux peaks at energy $\sim 10^{16}$~eV, and the peak energy flux is~\eFenu$\sim 3 \times 10^{-13}$~\ergcmsqs;  variations between the four EBL models are  $\leq 30\%$. The low-energy bump of the neutrino energy spectrum (at $\sim 10^{14}$~eV) is due to neutron decay. The expected neutrino rate in IceCube is  $10^{-3}~\mathrm{yr}^{-1}$ and $3 \times 10^{-4}~\mathrm{yr}^{-1}$ for the Point Source  and Alert searches, respectively, assuming the EBL model of~\citet{2012MNRAS.422.3189G}. The corresponding neutrino spectrum is also compared to those from the other scenarios we explored in Figure~\ref{fig:Aeff-specv}. 

In the IGC scenario, the interactions of the HECRs with the background photons produce not only neutrinos, but also $\gamma$-rays. These secondary $\gamma$-rays contribute additional energy flux in the GeV-TeV energy range of the SED of the source. In the example of Figure~\ref{fig:igc}, the maximum proton energy was chosen so as to be compatible with the parameters derived from the leptohadronic modeling, but also be below the threshold energy for photopair production on Cosmic Microwave Background (CMB) photons. 
Therefore, $\gamma$-rays and neutrinos are produced predominantly in interactions with the more energetic optical and infrared background photons with comparable energy flux channeled to the two messengers. As a result, in the example of Figure~\ref{fig:igc} the IGC $\gamma$-ray flux is well below the total $\gamma$-ray flux of \hsp\, inferred from the \fermi\, long-term observations, even if the strength of extragalactic magnetic fields is negligible (not explicitly shown). This is also due to our chosen value of the proton luminosity, $L_{p,esc}$. A much higher value of $L_{p,esc}$ would lead to a higher neutrino luminosity but also a higher $\gamma$-ray luminosity, possibly in conflict with the \fermi~spectrum of \hsp. 

A higher proton maximum energy, $\varepsilon_{p,\max} \gtrsim 10^{20}~\mathrm{eV} (\varepsilon_{\mathrm{CMB},z}/6\times 10^{-4} ~{\rm eV})$, where $\varepsilon_{\mathrm{CMB},z}$ is the characteristic energy of CMB photons at redshift $z$, would additionally allow neutrino and $\gamma$-ray production in interactions with CMB photons, thus increasing the expected neutrino and $\gamma$-ray energy flux. However, the neutrino flux produced in CMB interactions would peak at $\gtrsim$~EeV energy, owing to the high proton threshold energy. As shown in Figure~\ref{fig:Aeff-specv}, such high-energy neutrinos do not help explain \icv, owing to the smaller IceCube effective area at this declination. 

We also note that the proton cutoff energy used in the leptohadronic models of \hsp~(see Models A-D and HEP scenario) is typically much lower than the energy range of ultrahigh-energy cosmic rays (i.e., $>10^{18}$~eV). On the contrary, the IGC and PS models, which require much higher proton energies, are consistent with scenarios relating HBL with ultrahigh-energy cosmic rays~\citep[see][and references therein]{2012ApJ...749...63M}.

\begin{deluxetable}{cccc}
\centering
\tablecaption{Yearly rate of muon and antimuon neutrinos expected to be detected by IceCube, with the Point Source (PS) and Alert searches, for the alternative neutrino emission models of \hsp.
\label{tab:nu_rate_other}}
\tablewidth{0pt}
 \tablehead{
 \colhead{Model} & \colhead{State} &  $\dot{\mathcal{N}}_{\nu_{\mu}+\bar{\nu}_{\mu}}(> 100 ~\rm TeV)$ & $\mathcal{P}|_{1\,\nu_{\mu}\,{\rm or}\,\bar{\nu}_{\mu}}$ \\ &  & ($\times10^{-4}$ yr$^{-1}$) & $(> 100 ~\rm TeV)$ \\
 & &  Alert (PS) & Alert (PS) \\
 }
 \startdata
HEP & transient high & 50 (190) &  0.3~(1)\%\\
PS  & transient high & 2.1 (7.3) &  0.01~(0.05)\%\\
\hline
BC  & persistent average & 33 (370) & 3~(30)\% \\
IGC & persistent average & 3.6 (10) & 0.4~(1)\% \\ 
\enddata
\tablecomments{For the IGC scenario, we report the rate computed using the EBL model of~\citet{2012MNRAS.422.3189G}. For each model, we report whether the quoted rate corresponds to the persistent average emission state or to a transient high state based on the 250 day \fermi~high-state in 2019-20. In the rightmost column, we report the Poisson probability to detect one muon (or antimuon) neutrino with energy exceeding 100 TeV in 10 years and 250 days of IceCube livetime for models of the persistent average and transient-high emission, respectively.}
\end{deluxetable}

Summarizing, the rate of neutrinos expected to be detected with IceCube with all the models explored in this section is presented in Table~\ref{tab:nu_rate_other}. We find that the HEP and BC models, which are effectively multi-zone models\footnote{HEP can also work as an one-zone model, but with lower predicted neutrino flux than its multi-zone version (see Section~\ref{sec:hep}).}, result in significantly higher expected neutrino rates than the PS and IGC models. Note however, that the PS and HEP models describe the enhanced \fermi\, state of \hsp\, in 2019, (starting on MJD 58605.6) whereas the IGC and BC models are compatible with the long-term SED of the source, therefore a direct comparison is not possible. All in all, we find that the BC and HEP models predict a neutrino rate comparable to that of the leptohadronic models presented in Section~\ref{sec:modeling}, whereas the PS and IGC models predict a lower neutrino rate (compare with rates in Table~\ref{tab:nu_rate}).

\section{Discussion}\label{sec:discussion}
In this section  we present our results on the source energetics (Section~\ref{sec:jetpower}),  baryon loading factor, and  neutrino-to-$\gamma$-ray luminosity ratio (Section~\ref{sec:baryon}) as inferred by the single-zone leptohadronic models presented in Section~\ref{sec:modeling}. We also compare our findings with previously published results for BL Lac sources and  \txs, obtained in the framework of one-zone emission models.  We finally discuss the implications of our results for the high-energy neutrino \icv~ (Section~\ref{sec:neutrino}).

\subsection{Jet power}\label{sec:jetpower}
We comment on the energetic requirements of the standard one-zone leptohadronic models presented in Section~\ref{sec:results}. For each model, we compute the absolute power of a two-sided jet, as $P_{j}\approx (8\pi/3) R^{\prime 2} c \Gamma^2 \left(u^{\prime}_e+u^{\prime}_p+u^\prime_B \right)$ where $\Gamma \approx \delta$ and $u^\prime_e \ll u^\prime_{B}, u^\prime_p$ \citep[e.g.,][]{Zdziarski:2015rsa, PD16}. We then compare the derived $P_j$ values (see Table~\ref{tab:param-2}) to two characteristic  energy estimators of an accreting black-hole system, namely the Eddington luminosity, $L_{\rm Edd}$, and the power of the Blandford-Znajek (BZ) process, $P_{\rm BZ}$ \citep{Blandford1977}. 

Using an estimate for the black-hole mass, i.e., $M_{\rm BH}\sim 3\times10^8 M_{\odot}$ \citep{2020MNRAS.495L.108P},
we find $L_{\rm Edd}\sim 4\times10^{46}$~erg s$^{-1}$. The magnetic field threading the black-hole horizon is one of the usually unknown parameters needed to compute the BZ power of a spinning black hole \citep[e.g.,][]{Tchekhovskoy2011}. It can be inferred by radio-core shift measurements at large scales under certain assumptions \citep{Lobanov1998, Zdziarski2015-II,Finke2019}. For \hsp, however, this information is not available. We therefore compare our results with the BZ power of the  blazar sample studied in \cite{LP20}.
 
These authors computed $P_{\rm BZ}$ using the core-shift measurements for 47 blazars (composed of LBL and IBL sources), assuming that all sources host maximally spinning black holes. They also estimated the BZ power for 137 blazars  without core-shift measurements using the sample's median (and standard deviation) opening angle and magnetic field strength at 1~parsec \citep[for details on the derivation, we refer the reader to][]{LP20}.

\begin{figure}
    \centering
    \includegraphics[width=0.47\textwidth]{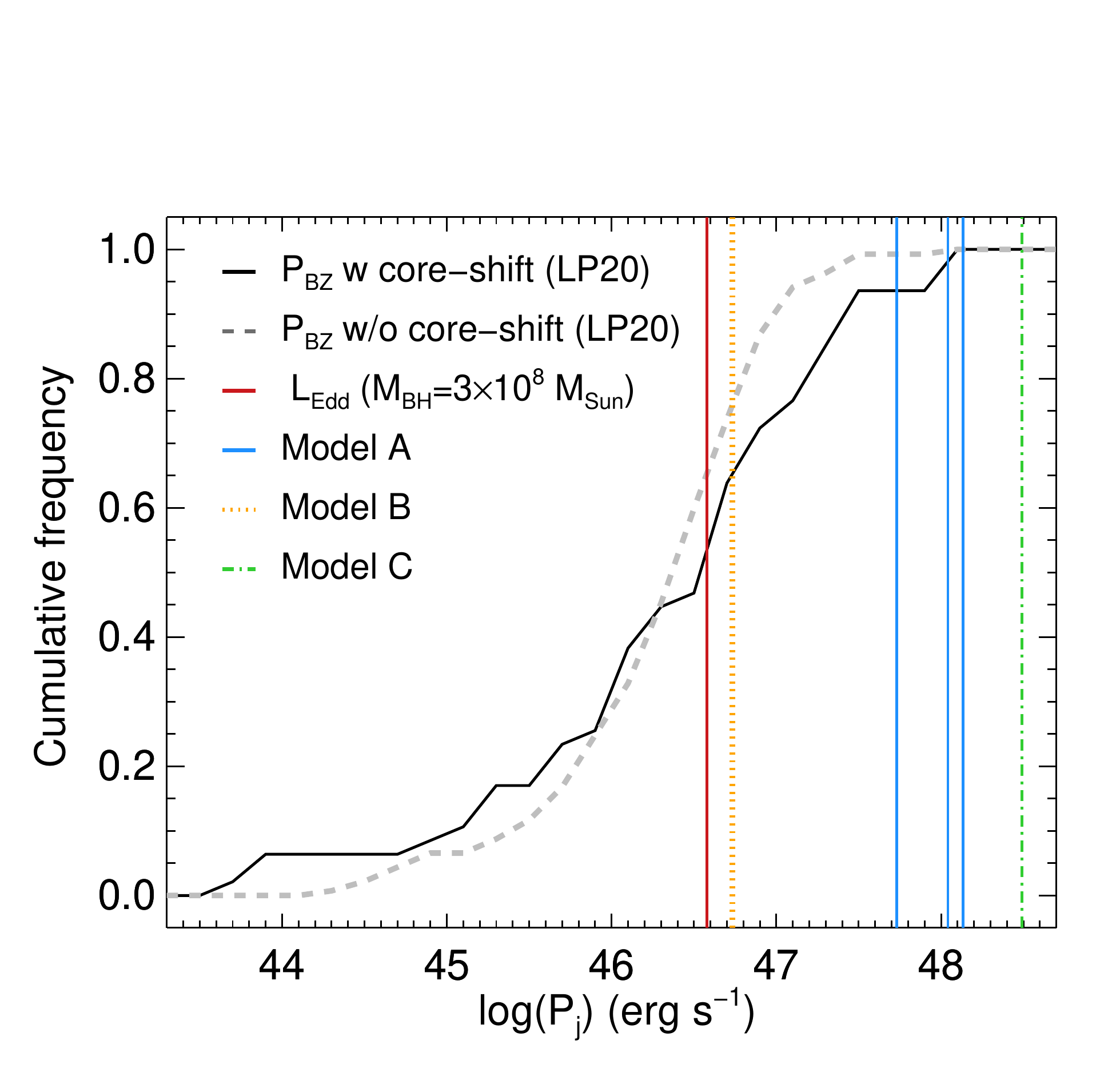}
    \caption{Cumulative distribution of $P_{\rm BZ}$ (in logarithmic units) for blazars with (solid black line) and without (dashed grey line) core-shift measurements \citep[adopted from][LP20]{LP20}. The vertical solid red line indicates the estimated Eddington luminosity of \hsp, and the remaining  vertical colored lines mark the jet power in leptohadronic Models A-C (see inset legend). Model D with $P_j\sim 6\times 10^{50}$~erg s$^{-1}$ is not shown.}
    \label{fig:pj-histo}
\end{figure}

Figure~\ref{fig:pj-histo} shows the cumulative distribution of $P_{\rm BZ}$ (in logarithmic units) for blazars with (solid black line) and without (dashed grey line) core-shift measurements. The vertical solid red line indicates the estimated Eddington luminosity of \hsp, and the remaining  vertical colored lines mark the jet power of the leptohadronic Models A-C discussed in Section~\ref{sec:results} (see inset legend). Model D is not shown in the figure, as it has extremely high jet power and falls well beyond the plotting range (see Table~\ref{tab:param-2}). Out of the remaining models, Model C is the most energetically demanding, exceeding $L_{\rm Edd}$ by $\sim 2$ orders of magnitude. Most importantly, the inferred jet power is higher than the maximum power of the BZ process found for the blazar sample of \cite{LP20}. Model C and, more generally, models with similarly low photomeson production efficiencies (see also Figure~\ref{fig:heatmap}), are therefore strongly disfavored at least for the long-term blazar emission. The $P_j$ values of Model A lie at the high-end of the BZ power distribution ($\sim 16\%$ of $P_{\rm BZ}$ values are higher than $P_{j}$ for Model A with $B^\prime=15$~G). Model B, which was selected to have the highest photomeson production efficiency of the three models, is the most plausible energetically, with $P_j$ close to the median of the $P_{\rm BZ}$ distribution and $P_j \sim L_{\rm Edd}$.

In general, the jet power in an accreting system can be written as $P_j=\eta_j \dot{M} c^2$, where $\dot{M}$ is the accretion rate onto the black hole and $\eta_j$ is the jet-formation efficiency, which can be as high as $\sim 1.5$ for magnetically arrested accretion  discs \citep[MAD,][]{Bisnovatyi1974, Narayan2003, Tchekhovskoy2011, Tchekhovskoy2012}. Using the upper limit on the bolometric accretion luminosity of \hsp, $L/L_{\rm Edd}<0.02$ and $L_d \sim 0.5 L$ \citep{Giommi2020}, we find $\dot{M} c^2 \lesssim (0.01/\epsilon) L_{\rm Edd}$, where $\epsilon<1$ is the radiative efficiency of the disc. This translates to $P_{j} \lesssim 6\times  10^{45}\, (\eta_j/1.5)(0.1/\epsilon)$~erg s$^{-1}$, assuming that the accretion happens in the MAD regime. All models studied here, except for Model B, require much higher jet powers than $10^{46}$~erg s$^{-1}$, and are therefore disfavored (at least for the average emission of \hsp). 

A more conservative upper limit on the accretion power can be derived if one adopts a different scaling relation between disk luminosity and accretion rate, $L_d \propto \dot{M}^2$ \citep{1995ApJ...452..710N, 1997ApJ...478L..79N}, that is more appropriate for low-excitation galaxies (LEGs) \citep[e.g.,][]{2014MNRAS.445...81S}, which is likely the case for \hsp~\citep{Giommi2020}. Using the upper limit $L/L_{\rm Edd}<0.02$, $L_d \sim 0.5 L$, and  $L_d/L_{\rm Edd}\sim \dot{m}^2/\dot{m}_{cr}$, we find $\dot{M}c^2 \lesssim 0.03\,  L_{\rm Edd} \left(\dot{m}_{cr}/0.1\right)^{1/2}$, where $\dot{m}\equiv \dot{M}c^2/L_{\rm Edd}$ and $\dot{m}_{cr}$ is a critical value of the accretion rate that separates different regimes of accretion \citep[e.g.,][]{1997ApJ...478L..79N}. In this case, the discrepancy between the maximum jet power (in MAD) and the model-predicted jet power would  be even larger.

\subsection{Baryon loading factor and neutrino-to-$\gamma$-ray luminosity ratio}\label{sec:baryon}
\begin{figure}
    \centering
    \includegraphics[width=0.47\textwidth]{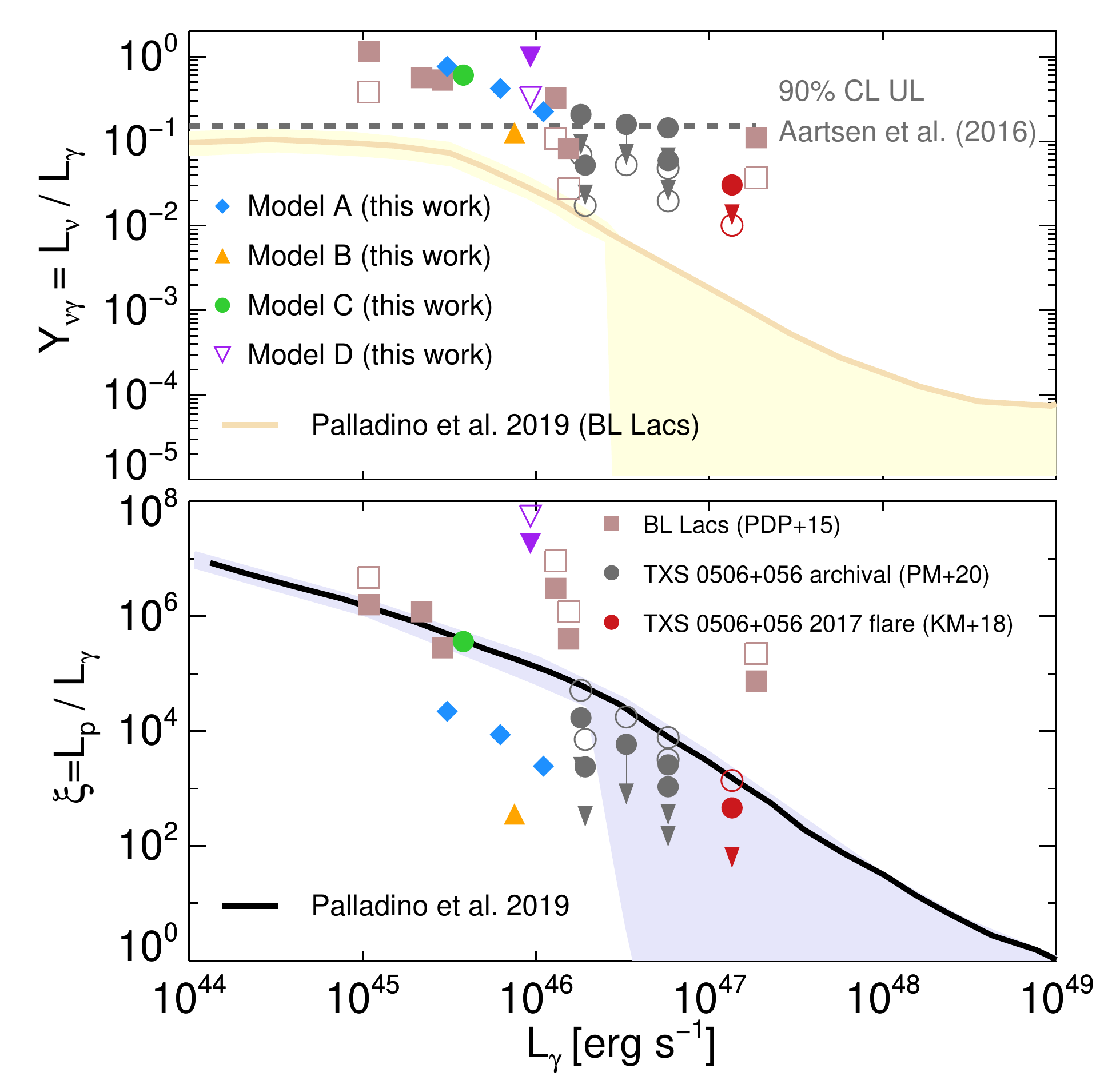}
    \caption{Baryon loading factor $\xi$ (bottom panel) and ratio of the neutrino to $\gamma$-ray luminosity $Y_{\nu \gamma}$ (top panel) of \hsp \, as a function of the $\gamma$-ray luminosity in the 0.1--300 GeV for Models A-D (see inset legend and Table~\ref{tab:param-2}). For comparison, we also show the maximum values of $\xi$ and $Y_{\nu \gamma}$ obtained for \txs, during its 2017 flare  \citep{Keivani2018} and for archival data \citep{Petropoulou2020}. Results for other six BL Lac objects from \cite{Petropoulou:2015upa} are also plotted. Filled and open symbols correspond to $s_p=1-1.3$ and $s_p=2$, respectively. The horizontal dashed line (top panel) marks the 90\% upper limit on $Y_{\nu \gamma}$ for the model of \cite{Padovani:2015mba} set by IceCube \citep{Aartsen:2016prl}. For illustration purposes, we overplot the baryon loading factor and $Y_{\nu \gamma}$ parameter with their uncertainty (shaded regions) from a model for the diffuse astrophysical neutrino flux at energies $\gtrsim 1$~PeV from blazars \citep[see scenario 3 in][]{palladino_2019}.}
    \label{fig:baryon}
\end{figure}

From the SED modeling, we can determine the baryon loading factor, defined as $\xi\equiv L_p/L_{\gamma}$, where $L_p = \dop^4 L^\prime_p$ is the isotropic-equivalent proton luminosity in the observer's frame and $L_{\gamma}$ is the $\gamma$-ray luminosity of the model in the $0.1-300$ GeV energy band. The neutrino luminosity of a blazar is commonly parameterized as $L_{\nu+\bar{\nu}}=Y_{\nu\gamma} \, L_{\gamma}$, where $L_{\nu + \bar{\nu}}$ is the all-flavor neutrino flux in the $0.1-10$ PeV energy range. $Y_{\nu\gamma}$, the neutrino-to-$\gamma$-ray luminosity ratio,  encodes information about the baryon loading and the neutrino production efficiency of the source~\citep{Petropoulou:2015upa, Padovani:2015mba, palladino_2019}. Roughly speaking,  $Y_{\nu\gamma}\sim(3/8)\fpg\xi$, where $\fpg$ is the photomeson production efficiency. Our results on $\xi$ and $Y_{\nu \gamma}$ for the models discussed in Section~\ref{sec:results} are summarized in Table~\ref{tab:param-2} and displayed in  Figure~\ref{fig:baryon}.

To put our findings into context, we complement  Figure~\ref{fig:baryon} with $\xi$ and $Y_{\nu \gamma}$ values obtained from the SED modeling of \txs \, during its 2017 multi-wavelength flare \citep{Keivani2018} and in four epochs prior to it \citep{Petropoulou2020}. The reported values for \txs \, are in fact upper limits (denoted by arrows in the figure), as its SED was modeled with processes of primary electrons alone, while the hadronic component was radiatively sub-dominant \citep[e.g.,][]{Ansoldi2018, Keivani2018, Gao2019}. In contrast to \txs, 
the SED of \hsp \, can be described by leptohadronic models, as it allows for a non-negligible contribution of secondary pairs to its high-energy  emission (see Figures~\ref{fig:sed}, \ref{fig:sed-2} and \ref{fig:sed-3}). In Figure~\ref{fig:baryon} we also include  the values derived by the SED modeling of six BL Lac objects \citep{Petropoulou:2015upa} that were identified as possible high-energy neutrino candidate sources by \cite{Padovani:2014bha}. We do not directly report the values listed in Table~2 of \cite{Petropoulou:2015upa}, because they were computed using different energy ranges for $L_{\gamma}$ and $L_{\nu + \bar{\nu}}$ than here. For consistency, we include in Figure~\ref{fig:baryon} updated values of $\xi$ and $Y_{\nu \gamma}$, computed using the same energy ranges for the luminosities as here.

The power-law slope of the proton distribution is usually unconstrained in leptohadronic models of blazar emission. Here we adopt for simplicity the same power-law index for the primary electron and proton distributions at injection, i.e.,  $s_p=1-1.2$ for Models A-C and $s_p=2$ for Model D (see Table~\ref{tab:param}). The upper limits on  $\xi$ and $Y_{\nu \gamma}$ reported in \citet{Keivani2018} and \citet{Petropoulou2020} were derived for the default choice of $s_p=2$. The same index was adopted for four out of the six blazars modeled by \citet{Petropoulou:2015upa}, while $s_p<1.3$ was used for the remaining sources. Both $\xi$ and $Y_{\nu \gamma}$ depend, however, on $s_p$. For a fixed target photon field, the flux of neutrinos (and other secondaries) produced via photomeson production increases with decreasing $s_p<2$, since more power is carried by protons with higher energies relevant for neutrino production \citep[see e.g., Figure 12 in][]{DMPR12}. Meanwhile, harder proton energy spectra (i.e., $s_p<2$) tend to decrease the required proton luminosity. 
More specifically, for $s_p\sim 1-1.2$ the neutrino luminosity can be $\sim 3$ times higher than the value derived for $s_p=2$, while the proton luminosity can decrease accordingly by a factor of $\sim 3$ \citep[see Figure~5 in][]{Petropoulou2020}. The original values (upper limits) obtained for $s_p=2$ are displayed in Figure~\ref{fig:baryon} as open squares (circles). Filled squares (circles) indicate the expected values (upper limits) for $\xi$ and $Y_{\nu \gamma}$, if $s_p \sim 1-1.2$. 
 
There is an emerging trend that $Y_{\nu \gamma}$ decreases with increasing $L_{\gamma}$. In other words, the contribution of secondaries from photomeson interactions to the high-energy blazar emission is smaller in sources that are more $\gamma$-ray luminous. Interestingly, the upper limits derived for \txs \, (after correcting for the different power-law slope of the proton energy spectrum used therein) seem to fall on the extension of a  line passing through the $Y_{\nu \gamma}$ values of \hsp. This trend is also supported by the luminosity ratios derived for six other BL Lac objects, characterized by different average $\gamma$-ray luminosities. Using these results, \citet{Petropoulou:2015upa} also reported on a tentative anti-correlation between $Y_{\nu \gamma}$ and $L_{\gamma}$, but because of the limited sample size this relation could not be confirmed at the time. 

The dependence of $Y_{\nu \gamma}$ on $L_{\gamma}$ is particularly important for models of the diffuse neutrino flux from the blazar population.  \citet{Padovani:2015mba} computed the contribution of BL Lac objects to the diffuse neutrino flux assuming a common value ($Y_{\nu \gamma}=0.8$) for all sources, since at the time there was no strong evidence for an anti-correlation between $Y_{\nu \gamma}$ and $L_{\gamma}$. IceCube upper limits on the diffuse neutrino flux at extremely high energies ($>1$~PeV) constrain the luminosity ratio to be $Y_{\nu \gamma} \lesssim 0.15$~\citep{Aartsen:2016prl} (see horizontal dashed line); the latest upper limits from IceCube push the limit to $\lesssim 0.1$~\citep{Aartsen:2018PhRvD}. Given that these upper limits apply in a scenario where  $Y_{\nu \gamma}$ is universal among BL Lac sources,  it is not alarming that the ratios derived for \hsp \, (and other individual sources) lie above that limit.  As another example, we show the  hypothetical relation between $Y_{\nu \gamma}$ and $L_{\gamma}$ adopted by \citet{palladino_2019} when modeling the diffuse neutrino flux from BL Lac objects (yellow line and shaded region). 

Despite the different source conditions of the models we studied here, there is  small scatter in the predicted neutrino-to-$\gamma$-ray luminosity ratios. Contrary to the $Y_{\nu \gamma}$ parameter, the baryon loading factor varies by orders of magnitude, as shown in the bottom panel of the figure. This is expected, since Models A-D have been selected to have different photomeson production efficiencies (see Figure~\ref{fig:heatmap}). Model D, which is the most inefficient in terms of photomeson production, requires an extremely high proton luminosity to account for similar $\gamma$-ray (and neutrino) luminosity as the other models. These results highlight the effect that the source parameters, such as size and Doppler factor, have on the baryon loading factor. Similar conclusions can be drawn also for the other BL Lac sources from \citet{Petropoulou:2015upa}. We note that the upper limits on $\xi$ for \txs \, were derived by modeling different epochs using the same source parameters. This explains the small scatter in the  maximum values of $\xi$ for \txs. So far, our results cannot reveal the intrinsic relation between the baryon loading factor and $\gamma$-ray luminosity, if any. Moreover, there no physically motivated scenario that predicts a negative correlation between $\xi$ and $L_{\gamma}$. Therefore, results of diffuse neutrino emission models from blazars that rely on such relations, as shown with the blue solid line in the bottom panel of the figure, should be considered with caution.

Summarizing, Figure~\ref{fig:baryon} highlights the importance of the SED modeling of individual blazars at different $\gamma$-ray luminosities (both during flares and epochs of low electromagnetic activity). With better multi-wavelength data availability for each source, in future, we will be able to draw more robust conclusions on a possible trend on $Y_{\nu\gamma}$ and eliminate any biases that might result from incomplete knowledge of the SED. Additionally, by populating such diagrams with more sources, we will be able to properly benchmark models for the diffuse neutrino emission and motivate theories to explain the observed trends.  

\subsection{Implications for \icv}\label{sec:neutrino}
We now discuss the implications of our modeling results for interpreting the putative association of \icv\, and \hsp. We found, from the modeling of the X-ray high-state in Section~\ref{sec:modeling}, that the maximal expected number of neutrinos during the 44 day period starting on January 8th 2020 is $N_{\nu_{\mu}+\bar{\nu}_{\mu}}(> 100~{\rm TeV}) = 6\times10^{-4}$ in the IceCube Alert analysis (this expectation corresponds to Model A with $B^{\prime} = 30$~G). The probability to detect one or more neutrinos with this expected $N_{\nu_{\mu}+\bar{\nu}_{\mu}}(> 100 ~\rm TeV)$ is low, i.e., $\sim 0.06\%$.  The expected number of neutrinos could be greater if the high X-ray state lasted for several years prior to the arrival of \icv, or if the X-ray flare reached its peak intensity before the first \swift \, observation on January 8, 2020. However, both possibilities remain highly speculative due to the lack of X-ray observations prior to January 8, 2020 since December 2013. 

We considered several scenarios for the long-term neutrino emission of \hsp\, in Section~\ref{sec:longterm}. The maximal expected number of neutrinos during the ten year period starting in April 2008 which marks the beginning of IceCube operation is  $N_{\nu_{\mu}+\bar{\nu}_{\mu}}(> 100 ~\rm TeV) = 0.01$  (see Figure~\ref{fig:neutrino_counts}). With this expectation value, the probability to see one or more neutrinos in ten years is $\sim1\%$.

In addition to the one-zone leptohadronic models explored in Section~\ref{sec:longterm}, we investigated alternate models in Section~\ref{sec:other}. Of those, the two most promising models in terms of neutrino production, and with comparable expected neutrino rates were found to be the blazar core (BC) model which considers neutrino production in the vicinity of the accreting supermassive black hole, and the hidden extrernal-photon (HEP) scenario which considers neutrino production through interactions with photons from a possible weak broad line region. These two models are effectively multi-zone scenarios, although the latter could also describe the full SED of \hsp\, in a single-zone scenario (but at the cost of a lower expected neutrino flux). The HEP model we have considered is constrained by the 250 day \fermi\, high-state of \hsp. The timescale of the BC model is unconstrained by the observations of \hsp. If the conditions required to produce blazar-core emission existed during a long timescale, the expected number of neutrinos in ten years in the alert channel is $N_{\nu_{\mu}+\bar{\nu}_{\mu}}(> 100 ~\rm TeV) = 0.03$, a factor of three higher than the maximal expected neutrino rate from the models of Section~\ref{sec:modeling}. The IGC model is the only model we investigated in which neutrino production happens outside the jet, in the intergalactic medium. We found that at the declination of \hsp\, this scenario is expected to produce a modest neutrino rate (see also Table~\ref{tab:nu_rate_other}). We therefore conclude that for interpreting IC-200107A, models in which neutrino production takes place in the jet are more promising.

A proper comparison of \hsp \, and \txs \, as neutrino sources will be possible after a search has been performed with IceCube in the archival data in the direction of \hsp. The expected number of neutrinos in the archival search, which has a larger effective area than the Alert search, for \hsp\, in our more optimistic models is $\sim 0.03$ $\nu_\mu + \bar{\nu}_\mu$ per year and thus comparable to (though slightly lower than) that of TXS 0506+056 ($\sim 0.04-0.2$ $\nu_\mu + \bar{\nu}_\mu$ per year) as found by \citet{Petropoulou2020}. Since all the models we investigated predict less than one neutrino, additional neutrinos are not expected with the archival IceCube search. However, the archival search is interesting even in the case of no detection of additional neutrinos, which we expect, because it will allow to revise the neutrino flux calculation.

Though the rate of neutrinos expected from \hsp\, is $\ll 1$ in all the models we studied, it follows from Equation~\ref{eq:Nnu}, that if instead of a single source producing a flux $\phi_{\varepsilon_{\nu}}$ we consider a population of neutrino producing sources, for example all or a subset of HBL blazars, with individual neutrino fluxes $\phi_{\varepsilon_{\nu},i}$ producing a summed expectation of order $\sim$one neutrino in IceCube, then the flux requirements on each individual source $i$, in this case \hsp, are significantly reduced. There exist approximately 100 blazars in the sky with properties similar to \hsp~\citep{Giommi2020}. If they all produce a comparable neutrino flux, then the summed expectation could be of order one, which is consistent with the diffuse neutrino flux measurement to which the contribution from HBL is likely to be sub-dominant.  

In future, the IceCube-Gen2 detector~\citep{Aartsen:2014njl} will operate in concert with KM3NeT~\citep{Bagley:2009wwa} and other proposed/upcoming facilities in the Northern hemisphere~\citep{2018arXiv180810353B,agostini2020pacific}. Assuming that the IceCube-Gen2 detector will have effective area ten times larger than IceCube and KM3NET similar effective area to that of IceCube, the long-term neutrino emission of \hsp\, would be expected to result in the emission of $\sim$1-3 muon neutrinos in ten years above 100 TeV based on the most optimistic models we have studied (see the procedure outlined  in~\citealp{Oikonomou:2019djc} for details). Considering the ensemble of $\sim 100$ blazars listed in the 3HSP catalog with properties similar to those of \hsp, if neutrino production proceeds optimally in all these sources, for example with conditions similar to those illustrated by Model A, the stacked neutrino signal from the long-term emission of these sources should be easily detectable with these upcoming neutrino detectors, or otherwise the models we have studied will be strongly constrained. 

\section{Summary}
\label{sec:summary}
\hsp~is an extreme blazar with synchrotron peak frequency $\gtrsim 2$~keV that has been possibly associated with a high-energy neutrino, \icv. The latter was detected one day before the blazar was detected in a hard X-ray state. Motivated by this observation, we have performed a comprehensive study of the predicted neutrino emission from \hsp~during its recent X-ray flare, but also during the lifetime of IceCube observations. 

We focused on single-zone leptohadronic models, where the blazar electromagnetic and high-energy neutrino emissions originate from same region of the jet, but we also explored alternative scenarios. These include  a blazar-core (BC) model, which considers neutrino production in the inner jet close to the accreting supermassive black hole, a hidden external-photon (HEP) model, which considers neutrino production in the jet through interactions with photons from a possible weak broad line region,  a one-zone proton synchrotron (PS) emission model, where high-energy protons produce $\gamma$-rays  in the jet via synchrotron, and an intergalactic cascade (IGC) model, where neutrinos are produced in the intergalactic medium by interactions of a high-energy cosmic-ray beam escaping the blazar.

Although the association of \icv \, with the hard X-ray flare is likely coincidental, we find that there is a $\sim 1\%$ or $3\%$ probability of the neutrino coming from the long-term (ten year-long) emission of the source when considering the most promising one-zone leptohadronic model or the effectively multi-zone BC model, respectively. Interestingly, the most promising scenarios for neutrino production in \hsp \, predict strong attenuation of TeV $\gamma$-rays within the source, thus potentially differentiating strong neutrino emitters from the rest of the extreme blazar population with hard $\gamma$-ray spectra extending to TeV energies. Future neutrino detectors, like IceCube-Gen2, should be able to provide additional evidence of neutrino production in \hsp~ and the extreme blazar population in general.

\acknowledgments
The authors would like to thank the anonymous referee for a timely and constructive report.
The authors would also like to thank Dr. Dimitrakoudis for providing Figure~\ref{fig:sketch}, and Theo Glauch for providing the \fermi-LAT data from the analysis of~\citep{Giommi2020}. M.P. acknowledges support from the Lyman Jr.~Spitzer Postdoctoral Fellowship and NASA Fermi grant No.~80NSSC18K1745. The work of K.M. is supported by the Alfred P. Sloan Foundation, NSF Grant No.~AST-1908689, and KAKENHI No.~20H01901. The work of F.O. is supported by the Deutsche Forschungsgemeinschaft through grant SFB\,1258 ``Neutrinos and Dark Matter in Astro and Particle Physics''. GV is supported by NASA Grant Number  80NSSC20K0803, in response to XMM-Newton AO-18 Guest Observer Program.


\bibliography{blazar}{}
\bibliographystyle{aasjournal}

\end{document}